\documentclass{article}
\usepackage{bm,latexsym,amsmath,amssymb,amsfonts,fancyhdr,color,graphicx,multirow,slashed,cite}
\usepackage[a4paper,bottom=3cm,top=2.5cm,head=0mm,width=17cm,dvipdfm]{geometry}
\usepackage[usenames,dvipsnames,svgnames,table]{xcolor}
\usepackage[colorlinks=true,
            linkcolor=blue,
            urlcolor=blue,
            citecolor=green,          
						bookmarks=true,
						bookmarksnumbered=true,
						breaklinks=true,
						pdfpagemode=Fullscreen,
						pdfstartview=FitBH]{hyperref}
\usepackage[dotinlabels]{titletoc}
\usepackage{titlesec}
\usepackage{authblk,ulem}

\numberwithin{equation}{section}
\allowdisplaybreaks[4]

\titlelabel{\thetitle.\quad \hspace{-0.8em}}
\titlecontents{section}
              [1.5em]
              {\vspace{4mm} \large \bf}
              {\contentslabel{1em}}
              {\hspace*{-1em}}
              {\titlerule*[.5pc]{.}\contentspage}
\titlecontents{subsection}
              [3.5em]
              {\vspace{2mm}}
              {\contentslabel{1.8em}}
              {\hspace*{.3em}}
              {\titlerule*[.5pc]{.}\contentspage}
\titlecontents{subsubsection}
              [5.5em]
              {\vspace{2mm}}
              {\contentslabel{2.5em}}
              {\hspace*{.3em}}
              {\titlerule*[.5pc]{.}\contentspage}

\newcommand{\titledef}{
Complementary Search of Fermionic Absorption Operators at Hadron Collider and Direct Detection Experiments
} 

\hypersetup{ pdfauthor = {Shao-Feng Ge},
	     pdftitle = {\titledef}, 
	     pdfsubject = {}, 
             pdfkeywords = {}, 
	     pdfcreator = {LaTeX with hyperref package},
	     pdfproducer = {dvips + ps2pdf} }

\definecolor{gesfblack}{rgb}{0,0,0}

\definecolor{gesfblue}{rgb}{0.08,0.42,0.76}

\definecolor{gesfgreen}{rgb}{0,1,0}

\definecolor{gesfgrey}{rgb}{0.5,0.5,0.5}

\definecolor{gesflanse}{rgb}{0.00,0.50,0.50}

\definecolor{gesfpurple}{rgb}{0.47,0.19,0.42}

\definecolor{gesfred}{rgb}{1,0,0}

\definecolor{gesfwhite}{rgb}{1,1,1}

\definecolor{gesfyellow}{rgb}{0.7,0.4,0.3}

\newcommand{\gsec}[1]{{\hypersetup{linkcolor=red}Sec.\,\ref{#1}\hypersetup{linkcolor=blue}}}
\newcommand{\gapp}[1]{{\hypersetup{linkcolor=red}App.\,\ref{#1}\hypersetup{linkcolor=blue}}}
\newcommand{\geqn}[1]{\hypersetup{linkcolor=blue}Eq.\,(\ref{#1})\hypersetup{linkcolor=blue}}
\newcommand{\gfig}[1]{{\hypersetup{linkcolor=violet}Fig.\,\ref{#1}\hypersetup{linkcolor=blue}}}
\newcommand{\gtab}[1]{{\hypersetup{linkcolor=gesflanse}Table~\ref{#1}\hypersetup{linkcolor=blue}}}

\definecolor{Orange}{cmyk}{0,0.61,0.87,0}
\definecolor{JungleGreen}{cmyk}{0.99,0,0.52,0}
\definecolor{OliveGreen}{cmyk}{0.64,0,0.95,0.40}
\definecolor{Brown}{cmyk}{0,0.81,1,0.60}
\definecolor{RoyalBlue}{cmyk}{0.71,0.53,0,0.12}
\definecolor{Gray}{cmyk}{0,0,0,0.40}
\definecolor{LightPink}{cmyk}{0.0,0.25,0,0}
\definecolor{LLightPink}{cmyk}{0.0,0.10,0,0}
\definecolor{LightBlue}{cmyk}{0.25,0,0,0}
\definecolor{LightGray}{cmyk}{0,0,0,0.2}

\graphicspath{{figs/}}

\usepackage{subfigure}
\newcommand{\bee}{\begin{equation}}
\newcommand{\ene}{\end{equation}}
\newcommand{\bea}{\begin{eqnarray}}
\newcommand{\ena}{\end{eqnarray}}

\def\fb{\, {\rm fb}}

\def\tev{\,{\rm TeV}}
\def\gev{\,{\rm GeV}}

\def\call{\mathcal{L}}
\def\calm{\mathcal{M}}

\def\calo{\mathcal{O}}

\def\d{{\rm d}}

\def\fb{\, {\rm fb}}

\def\met{{E\!\!\!\slash}_{T} }

\def\ie{{\it i.e.}}
\def\etc{{\it etc.~}}

\def\iab{$\,{\rm ab}^{-1}$}
\begin{document}
\fontsize{12pt}{14pt}\selectfont

\title{
       \Large
       \bf \titledef} 
\author[1]{{\large Kai Ma} \footnote{Corresponding Author: \href{mailto:makai@ucas.ac.cn}{makai@ucas.ac.cn}}}
\affil[1]{Faculty of Science, Xi'an University of Architecture and Technology, Xi'an, 710055, China}
\author[2,3]{{\large Shao-Feng Ge} \footnote{Corresponding Author: \href{mailto:gesf@sjtu.edu.cn}{gesf@sjtu.edu.cn}}}
\affil[2]{Tsung-Dao Lee Institute \& School of Physics and Astronomy, Shanghai Jiao Tong University, Shanghai 200240, China}
\affil[3]{Key Laboratory for Particle Astrophysics and Cosmology (MOE) \& Shanghai Key Laboratory for Particle Physics and Cosmology, Shanghai Jiao Tong University, Shanghai 200240, China}
\author[4]{{\large Lin-Yun He}
\footnote{\href{mailto:a1164432527@gmail.com}{a1164432527@gmail.com}}}
\affil[4]{Center of Advanced Quantum Studies, School of Physics and Astronomy, 
Beijing Normal University, Beijing, 100875, China}
\author[2,3]{{\large Ning Zhou} \footnote{\href{mailto:nzhou@sjtu.edu.cn}{nzhou@sjtu.edu.cn}}}

\date{}

\maketitle

\vspace{-2mm}
\begin{abstract}
\fontsize{12pt}{14pt}\selectfont
Instead of the energy recoil signal at direct detection experiments, 
dark fermion appears as missing energy at hadron colliders. 
For a fermionc dark sector particle that coupled with quarks and neutrino
via absorption operators,
its production at collider is accompanied by an invisible neutrino.
We study in details the mono-$X$ (photon, jet, and $Z$) productions at the
Large Hadron Collider (LHC). We start from the quark-level absorption
operators to make easy comparison between the collider and direct detection
experiments. In other words, we study the model-independent
constraints on a dark fermion with absorption operator. In addition,
the interplay and comparison with the possible detection at the neutrino
experiments, especially Borexino, is also briefly discussed. We find that
light nuclear target can provide the stronger constraints on
both spin-dependent and spin-independent absorption operators.
\end{abstract}


\newpage

\section{Introduction}
\label{sec:intro}
Both astrophysical and cosmological measurements show that dark matter (DM) 
is an important component of 
our Universe \cite{Bertone:2004pz,Young:2016ala,Arbey:2021gdg,Fairbairn:2022gar}.
However, so far we are still short of direct observations
which are necessary clues for the DM model building 
\cite{Roszkowski:2017nbc,deSalas:2020hbh,Chadha-Day:2021szb,Heinemeyer:2022anz,Capolupo:2023wri,Arguelles:2023nlh}.
A general principle is that the DM should be neutral, massive, 
and weakly coupled to the  Standard Model (SM) particles.
In addition, it should be stable enough to have the correct DM
relic abundance.
Such Weakly Interacting Massive Particle (WIMP) is the most popular DM candidate
and its phenomenology has been extensively investigated in both
theoretical and experimental aspects.
Usually, the DM particles emerge in pair as a consequence of some
discrete symmetry such as $\mathbb{Z}_2$ for guaranteeing the DM stability. 
However, as long as the DM is light enough, its decay can be slow
enough to survive until today as the genuine DM candidate
\cite{Dror:2019onn,Dror:2019dib,Dror:2020czw,Ge:2022ius,Ge:2023wye}.

On the other hand, it is also possible that the genuine DM is only 
a constituent of a much richer dark sector (DS) of particles \cite{Deliyergiyev:2015oxa,Marra:2019lyc,Hofmann:2020wvr,Lagouri:2022ier,Gori:2022vri}.
The DS particles can couple to the SM particles either directly or indirectly
through the so-called ``portal" interactions.
In either case, the DS-SM couplings 
are only mildly constrained as it is usually assumed that 
the corresponding interactions are weak.
In the effective filed theory (EFT) approach of the DS-SM interactions,
UV dependence of the constraints on the involved couplings
\cite{Deliyergiyev:2015oxa,Marra:2019lyc,Hofmann:2020wvr,Gori:2022vri,Lagouri:2022ier} can be released, and relatively stronger exclusion limits 
can be obtained
by indirect observations
\cite{Bouquet:1989sr,Baltz:2002we,Ibarra:2009bm,Bi:2009uj,Gaskins:2016cha,Leane:2020liq,Slatyer:2021qgc,deDiosZornoza:2021rgw,Das:2021hnk,John:2021ugy,deLaurentis:2022oqa,Hutten:2022hud,Brito:2022lmd,
Reis:2024wfy},
direct detections \cite{MarrodanUndagoitia:2015veg,Liu:2017drf,Schumann:2019eaa,Billard:2021uyg,DelNobile:2021wmp,Cebrian:2021mvb,Misiaszek:2023sxe},
and collider searches 
\cite{Penning:2017tmb,Kahlhoefer:2017dnp,Boveia:2018yeb,Lorenz:2019pxf,Argyropoulos:2021sav,Ilten:2022lfq,Krnjaic:2022ozp}.
Furthermore, in the EFT framework, the interaction operators 
which are responsible to the low energy couplings between the DS particles and atom 
can also be explored at high energy colliders due to the fundamental interactions 
between quarks/leptons and DS particles
\cite{Dreiner:2013vla,Buchmueller:2014yoa,DEramo:2014nmf,Bertuzzo:2017lwt,Bishara:2017pfq,Belyaev:2018pqr,Cepedello:2023yao,Roy:2024ear}.
In this sense, the high energy collider searches and low energy direct detections can provide complementary constraints on the involved parameters.

With high center-of-mass energy, collider search has an additional superiority:
it can even probe the heavy DS particles that are usually contained in a UV model
\cite{Deliyergiyev:2015oxa,Marra:2019lyc,Hofmann:2020wvr,Gori:2022vri,Lagouri:2022ier},
as long as its direct production is kinematically allowed. 
The collider search can also provide a controllable environment to distinguish 
the leptophilic and hadrophilic nature of the DS couplings with SM
particles. 
While hadron colliders mainly probe quark and gluon couplings to DS particles
\cite{Su:2009fz,Farzan:2010mr,delAguila:2014soa,Buckley:2015cia,Ma:2024aoc}, 
lepton colliders are more sensitive to the leptonic ones 
\cite{Bartels:2012ex,Dreiner:2012xm,Freitas:2014jla,Habermehl:2020njb,Bharadwaj:2020aal,Kalinowski:2021tyr,Kundu:2021cmo,Barman:2021hhg,Liang:2021kgw,Ge:2023wye}. 
Furthermore, the isospin violation and lepton universality of the 
DS-quark and DS-lepton couplings can also be studied
at hadron and lepton colliders, respectively.
Therefore, collider search of the DS particles is not just complementary
but also unique for DS particles (or WIMP)
\cite{Bertuzzo:2017lwt,Bishara:2017pfq,Belyaev:2018pqr,Chakraborti:2021mbr,Boveia:2022adi,Alanne:2022eem,Gninenko:2023pkv,Ge:2023wye,Ma:2024aoc}.

In this paper, we focus on the four-fermion effective operators involving
a fermionic DS particle $\chi$ and quarks. While those effective operators with
a pair of DS particles
\cite{Dreiner:2013vla,Bertuzzo:2017lwt,Bishara:2017pfq,Belyaev:2018pqr,Cepedello:2023yao}
can induce elastic scattering off nuclei target
in direct detection experiments, they can also be probed at hadron colliders
via mono-photon \cite{Gershtein:2008bf,Gabrielli:2014oya,Abdallah:2015uba,daSilveira:2023hmt,Hicyilmaz:2023tnr}, mono-$Z/W$
\cite{Bell:2012rg,Alves:2015dya,Abdallah:2015uba,Bell:2015rdw,No:2015xqa,Yang:2017iqh,Wan:2018eaz,Abdallah:2019tpo,Kawamura:2023drb} and mono-jet process 
\cite{Abdallah:2015uba,Belyaev:2018ext,Bai:2015nfa,Belwal:2017nkw,Claude:2022rho}.
These mono-photon, mono-$Z/W$, and mono-jet channels are also known as mono-$X$ processes \cite{Liew:2016oon,Bernreuther:2018nat,Krovi:2018fdr,Bhattacharya:2022qck}.  
In this paper, we generalize such searches to the four-fermion absorption
operators formed by a fermionic DS particle $\chi$, 
a neutrino, and two quarks. 
Such interactions can induce the neutral-current DS particle absorption
at the nuclei target 
\cite{Dror:2019onn,Dror:2019dib,PandaX:2022osq,Majorana:2022gtu,CDEX:2022rxz,Li:2022kca,Ge:2024euk,Ma:2024aoc}.
At hadron colliders, the associated production of a DS particle $\chi$ 
and a neutrino carries away missing energy to induce the same final-state topology as the DS particle pair production in the mono-$X$ processes.
However, with different kinematic properties of the missing energy,  
constraints on the relevant parameters can also be significantly different.

The rest of this paper is organized as follows.
In \gsec{sec:eft}, we summarize the four-fermion contact interactions
and discuss the interplay of their detection signals 
at low- and high-energy experiments. 
The following \gsec{sec:Bound:LHC} studies the signal and (irreducible) 
background properties of the mono-$X$ production processes 
as well as the constraints from the current LHC data. The
details of the mono-photon, mono-jet, and mono-$Z$ productions are given 
in the subsections \gsec{sec:monoa}, \gsec{sec:monoj}, and \gsec{sec:monoz},
respectively. Then we discuss the projected sensitivities at
HL-LHC and HE-LHC in \gsec{sec:future}. In \gsec{sec:ncc}, we further
study the absorption process of a light DS paticle on a nuclei target.
Both the spin-independent and spin-dependent scatterings are considered
separately in \gsec{sec:DD:SI} and \gsec{sec:DD:SD}, respectively.
The possible interference effects of these two channels for the tensor operator
are investigated in \gsec{sec:DD:tensor} and
our conclusions can be found in \gsec{sec:conclusion}.

\section{Fermionic Dark Particle Absorption Operators}
\label{sec:eft}
 
In this paper, we study the interactions that can induce
the absorption of a fermionic DS particle by nuclei
\cite{Dror:2019onn,Dror:2019dib,PandaX:2022osq,Majorana:2022gtu,CDEX:2022rxz,Li:2022kca,Ge:2024euk,Ma:2024aoc}.
Instead of transferring the kinetic energy of a fermionic
DS particle into 
the nuclei recoil energy by elastic scattering, 
the fermionic DS particle is converted to a neutrino
in the absorption process. 
Both its mass and kinetic energy are transferred into 
the nuclei recoil and neutrino energies.
Such process can be effectively initiated by four-fermion
contact interactions
\cite{Dror:2019onn,Dror:2019dib,Ge:2024euk,Ma:2024aoc}
which is
very similar to the absorption process on an electron target
\cite{Dror:2020czw,Ge:2022ius,Li:2022kca,PandaX:2022ood,EXO-200:2022adi,Ge:2023wye,CDEX:2024bum}
.

In the effective field theory (EFT) framework, the interaction operators between
the DS and SM particles are usually constructed according to 
the SM $SU_C(3) \times SU_L(2) \times U_Y(1)$ gauge symmetries. 
Assuming that the isospin symmetry is preserved and denoting the first generation 
up ($u$) and down ($d$) quarks as an isospin doublet $q \equiv (u~d)^T$, 
the relevant leading order dim-6 operators connecting the quark
and DS paticle-neutrino current can be written as,
\begin{subequations}
\label{eq:effo}
\begin{align} 
  \mathcal{O}_{S}
& \equiv(\bar{q} q)\left(\bar{\nu}_{L} \chi_{R}\right) \,,
\label{eq:oS}
\\ 
  \mathcal{O}_{P}
& \equiv\left(\bar{q} i \gamma_{5} q\right)\left(\bar{\nu}_{L} \chi_{R}\right) \,,
\label{eq:oP}
\\ 
  \mathcal{O}_{V}
& \equiv\left(\bar{q} \gamma_{\mu} q\right)\left(\bar{\nu}_{L} \gamma^{\mu} \chi_{L}\right) \,,
\label{eq:oV}
\\ 
  \mathcal{O}_{A}
& \equiv\left(\bar{q} \gamma_{\mu} \gamma_{5} q\right)\left(\bar{\nu}_{L} \gamma^{\mu} \chi_{L}\right) \,,
\label{eq:oA}
\\ 
  \mathcal{O}_{T}
& \equiv\left(\bar{q} \sigma_{\mu \nu} q\right)\left(\bar{\nu}_{L} \sigma^{\mu \nu} \chi_{R}\right) \,,
\label{eq:oT}
\end{align} 
\label{eq:O}
\end{subequations}
as well as their hermitian conjugates. 
Although we put the left-handed neutrinos here,
it is equivalent to replace by its right-handed counterpart.
The quark currents $\bar{q} \varGamma q $ with $\varGamma \equiv 1$, $i\gamma_5$, $\gamma_\mu$, $\gamma_\mu\gamma_5$ and 
$\sigma_{\mu\nu}$ should be understood as
$\bar{q} \varGamma q = \bar{u} \varGamma u + \bar{d} \varGamma d$ 
from the isospin symmetry assumption. 
The above parameterization includes all the five independent Lorentz
structures of the quark bilinear and is complete in the sense that
any other dim-6 operator can be written as a linear combination of 
the above 5 operators by employing the $\gamma$ algebra and 
Fierz transformations \cite{Nieves:2003in,Nishi:2004st,Liao:2012uj}.
The neutrino is assumed to be left-handed while the dark matter $\chi$ is 
a Dirac fermion with both left- and right-handed components. 
The effective Lagrangian takes the form as,
\bee
\mathcal{L}_{\text {eff }}
=
\sum_{i} \frac{1}{\varLambda_{i}^{2}} \mathcal{O}_{i}+\text {h.c.},
\ene
with each operator having a cut-off scale $\Lambda_i$
of the possible fundamental physics.

The four-fermion contact interactions in \geqn{eq:effo} 
can lead to the associated production of DS particle and neutrino at hadron collider.
The above EFT description of collider searches can 
work well as long as 
the cut-off scale $\varLambda_i$ \cite{Dreiner:2013vla}
or the mediator mass 
\cite{Busoni:2013lha,Busoni:2014sya,Busoni:2014haa}
is much larger than the collision energy.
More concretely, 
suppose that couplings of the DS paticle $\chi$ and the quarks 
to a dark vector $V'$ of a dark $U(1)$ gauge group
are $g_\chi$ and $g_q$, respectively,
and the mixing strength of $\chi$ and
neutrino as induced by a dark scalar is $\varTheta$,
then one has $m_{V'} = \varTheta g_\chi g_q\varLambda$
for the heavy dark mediator $V'$ \cite{Dror:2019dib}.
Requiring $m_{V'} \gtrsim \varLambda$ implies $\varTheta g_\chi g_q \gtrsim 1$.
For a conservative estimation, 
we simply require $g_\chi \sim g_q \sim 4\pi$ for validation of perturbativity. 
Then, the mixing strength parameter can be much smaller,
$\varTheta \gtrsim 6.3\times10^{-3}$. Since such a
weak mixing effect have not been excluded experimentally,
our effective operators descriptions are still valid for
a conservative estimation of the exclusion limits.
Models with lighter mediator ($<\varLambda$) can get more stringent constraints.

Because of the running effect, 
the EFT operators defined at the TeV scale can become a mixture at low energy
\cite{Hill:2011be,Frandsen:2012db,Vecchi:2013iza,Crivellin:2014qxa,DEramo:2016gos,Bertuzzo:2017lwt,Bishara:2017pfq,Belyaev:2018pqr}.
However, the mixing effect of the effective operators can be numerically relevant 
only when the DS particle couples with the top quark
\cite{Bishara:2017pfq}.
At hadron colliders, the tree-level direct production of the DS particle is 
sensitive to only the light quarks. Here we assume that the effective operators in \geqn{eq:effo} 
is only valid for light quarks and hence the mixing effect can be safely neglected.
The interaction between the DS particle and top quark can be searched for via the associated production
of top quark(s) and DS particle at hadron collider while its influences on the running effect can
be investigated consistently. We will study this part elsewhere.

We would like to point out again that the DS particles need
not to be the DM that survives until now in the Universe.
Any DS fermion can have absorption operator with neutrino.
Although such operator can lead to decay of the DS fermion,
its stability is not a requirement if they are not DM.
For a dark fermion mass with \(m_\chi \lesssim 5\ \text{GeV}\),
a variety of thermal and non-thermal production mechanisms for such light dark fermions can accommodate the observed dark matter relic abundance \cite{Dror:2019onn,Dror:2019dib}. In this paper, rather than elaborating on these production mechanisms, we adopt a model-independent approach to study signal properties at hadron colliders and direct detection experiments. 
A collider search can probe not just DM but the DS particles
in general as detailed below in \gsec{sec:Bound:LHC}.
The complementary search at direct detection experiments
for a DM particle with small enough mass to guarantee its
stability can be found in \gsec{sec:ncc}.

\section{Constraints from LHC}
\label{sec:Bound:LHC}

\subsection{Mono-Photon }
\label{sec:monoa}
\begin{figure}[t]
\centering
\includegraphics[width=0.3\textwidth]{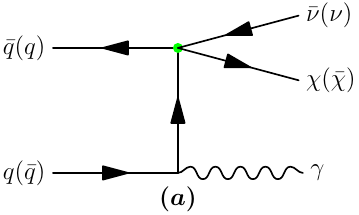}
\quad
\includegraphics[width=0.3\textwidth]{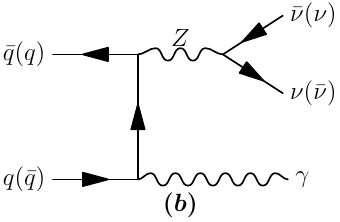}
\caption{\it 
Feynman diagrams of the mono-$\gamma$ process for
(a) the signal operators and (b) the irreducible background.
}
\label{fig:feyn:ma:LHC}
\end{figure}

The mono-photon radiation is one of the most powerful channel to search for dark particles 
at colliders \cite{Gershtein:2008bf,Gabrielli:2014oya,Abdallah:2015uba,daSilveira:2023hmt,Hicyilmaz:2023tnr}. The effective operators in \geqn{eq:effo} can induce
mono-$\gamma$ events through the initial-state radiation.
There are two subprocesses, $q\bar{q}\to \gamma\chi\bar{\nu}$ and 
$q\bar{q}\to \gamma\bar\chi \nu$ via the $t$-channel exchange of an off-shell quark,
that can contribute to the signal, as shown in \gfig{fig:feyn:ma:LHC}\,(a).
Since we have assumed that isospin is conserved in the effective absorptive interactions in \geqn{eq:effo}, the parton-level production rates depend only on the quark charges and 
masses. At LHC, the light quark masses can be safely neglected.
Hence, the signal cross sections at the parton level 
are universal for all the light quarks except for the charge dependence. The 2D differential cross sections are,
\begin{subequations}
\begin{align}
\label{eq:xs:sig:osp}
\frac{ d\hat\sigma_{S(P)}^q }{ dm_{X}^{2} d\cos\theta_\gamma }
&=
\frac{  Q^{2}_{q} }{128\pi^3 \varLambda^{4}_{S(P)} }
\frac{ (m_{X}^{2} - m_{\chi}^{2} )^2( \hat{s}^2 + m_{X}^{4}) }{ \hat{s}^2 ( \hat{s} - m_{X}^{2}) m_{X}^{2} \sin^2\theta_\gamma } \,,
\\[2mm]
\frac{ d\hat\sigma_{V(A)}^q }{ dm_{X}^{2} d\cos\theta_\gamma }
&=
\label{eq:xs:sig:ova}
\frac{ Q^{2}_{q}  }{768\pi^3 \varLambda^{4}_{V(A)} }
\frac{ (m_{X}^{2} - m_{\chi}^{2} )^2( 2 m_{X}^{2} + m_{\chi}^{2} ) \hat{s} }
{ ( \hat{s} - m_{X}^{2}) m_{X}^{6} \sin^2\theta_\gamma }
\left( f_{V(A)} + g_{V(A)} \cos\theta_\gamma \right) \,,~~
\\[2mm]
\label{eq:xs:sig:ot}
\frac{ d\hat\sigma_{T}^q }{ dm_{X}^{2} d\cos\theta_\gamma }
&=
\frac{ Q^{2}_{q}   }{1536\pi^3 \varLambda^{4}_{T} }
\frac{ (m_{X}^{2} - m_{\chi}^{2} )^2(  m_{X}^{2} +  2 m_{\chi}^{2}  ) \hat{s}^2}{( \hat{s} - m_{X}^{2}) m_{X}^{8}  \sin^2\theta_\gamma }
\left( f_{T} + g_{T} \cos\theta_\gamma \right) \,,
\end{align}
\label{eq:xs:sig}
\end{subequations}
where $\hat{s}$ is the center-of-mass energy at the parton level, 
$\theta_\gamma$ is the photon polar angle in the center-of-mass frame,  
and $m_{X}^{2} \equiv (p_\nu + p_\chi)^2$ is the DS particle-neutrino
invariant mass.
For convenience, we have defined the following functions
$f_{V,A,T}$ and $g_{V,A,T}$, 
\begin{subequations}
\begin{align}
  f_{V(A)}
& \equiv
  1 + 2 z_X + 10  z_X^2 + 2 z_X^3 + z_X^4,
\\[2mm]
  g_{V(A)}
& \equiv
  ( 1 - z_X )^2 (1 + 4 z_X + z_X^2 ),
\\[2mm]
  f_{T}
& \equiv
  17 -4 z_X + 7z_X^2 + 56 z_X^3 + 55 z_X^4 - 4z_X^5 + z_X^6,
\\[2mm]
  g_{T}
& \equiv
  (1- z_X)^3 (15 + 33 z_X + 15 z_X^2 + z_X^3),
\end{align}
\end{subequations}
in terms of $z_X \equiv m_X^2/\hat{s}$.

One can see that, there are collinear singularities encoded
as the factor $\sin\theta_\gamma$ in the denominators of
\geqn{eq:xs:sig} for all the signal operators. 
There are also soft singularities when the DS particle-neutrino invariant
mass approaches the center-of-mass energy, $m_X^2 \to s$.
Such singularities can be cured by cutting the photon transverse
momentum and the DS particle-neutrino invariant mass. The irreducible background
also has collinear and soft singularities that come from the
associated production of a photon and a $Z$ boson followed
by the invisible decay $Z\to\nu\bar\nu$.

The channels, $q\bar q' \to \gamma W^{\pm}(\ell^\pm\nu)$ and 
$q\bar q \to \gamma Z(\ell\ell)$ with the charged lepton(s)
not being detected 
as well as $q \bar q' (qg/gg) \to \gamma q \bar q'(qg/qq)$ with
the final-state quark(s)/gluon escaping the detector,
can also contribute to the total background. Fortunately,
they can be significantly reduced
by requiring large transverse missing energy. Hence, here we study
the kinematical distribution 
of the irreducible background, \ie, $q\bar q \to \gamma Z(\nu\nu)$.

\begin{figure}[t]
\centering
\includegraphics[width=0.32\textwidth]{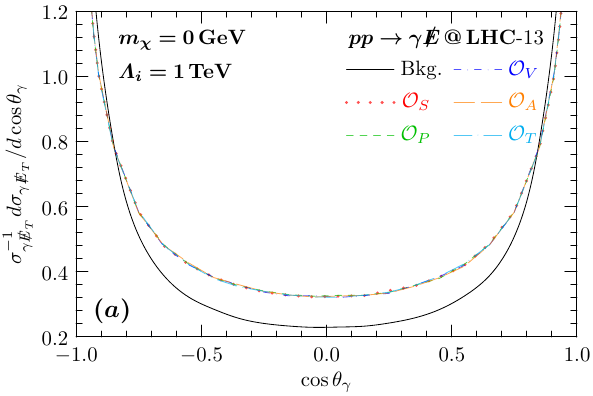}
\hfill
\includegraphics[width=0.32\textwidth]{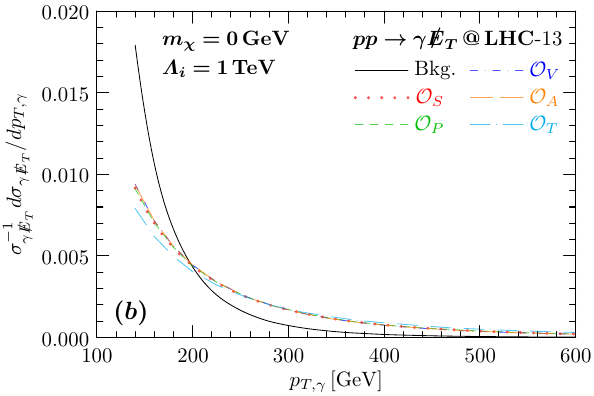}
\hfill
\includegraphics[width=0.32\textwidth]{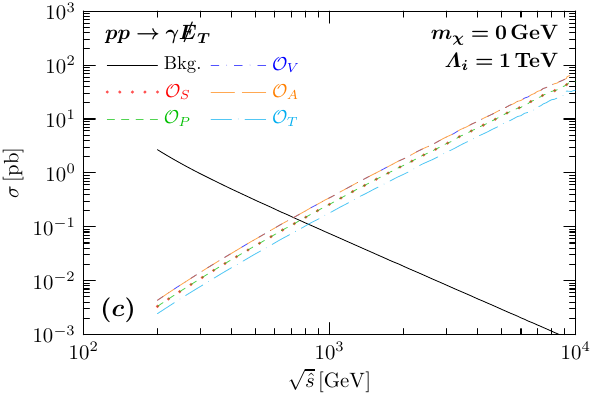}
\caption{\it 
The normalized parton-level distributions of the photon polar angle
($\theta_\gamma$) \textbf{(a)} and transverse momentum
($p_{T,\gamma}$) \textbf{(b)}
in the laboratory frame with the center-of-mass energy $\sqrt{s} = 13\tev$. 
The panel \textbf{(c)} is the total background and signal cross
sections as functions of the
center-of-mass energy $\sqrt{\hat{s}}$ at the parton level. 
In all the above panels, the signals (colorful non-solid curves)
are obtained with parameters $m_\chi = 0\gev$ and
$\varLambda_i = 1\tev$ while the background (black-solid curve)
stands for the irreducible
contribution from the channel $q\bar q \to \gamma Z(\nu\nu)$.
}
\label{fig:ca:pt:ma:LHC13}
\end{figure}
The panels (a) and (b) of \gfig{fig:ca:pt:ma:LHC13} show
the photon normalized polar angle $\theta_\gamma$ and
transverse momentum $p_{T,\gamma}$ distributions 
in the laboratory frame at LHC-13 ($\sqrt{s}=13\tev$
and a total luminosity $\call=139\fb^{-1}$).
Although the collider search can scan a much wider
mass range, we choose $m_\chi = 0\gev$ and $\varLambda_i = 1\tev$
to illustrate the signal properties
(colorful non-solid curves). For a small enough DM
mass, $m_\chi \ll \sqrt{\hat s}$ where $\sqrt{\hat s}$
is the effective center-of-mass energy in the primary
scattering process, the event cross section and topology
would become essentially insensitive to the DM mass.
Then, the illustrated results with vanishing mass can
apply more generally.
As expected, 
the signal events are dominant at the forward and backward regions
which is similar as the irreducible background distribution
(black-solid curve).
We can also see in \gfig{fig:ca:pt:ma:LHC13}\,(b) that
both the signal and background events mostly distribute
in the soft $p_{T,\gamma}$ region.
However, the signal distributions are larger than its background counterpart
for $p_{T,\gamma} \gtrsim 200\gev$. Hence $p_{T,\gamma} = 200\gev$ 
is a good cut to enhance the signal significance. 
\gfig{fig:ca:pt:ma:LHC13}\,(c) shows the dependence of 
the signal and background total parton-level cross sections
on the center-of-mass energy $\sqrt{\hat{s}}$.
While the background decreases quickly with increasing 
$\sqrt{\hat{s}}$, the signals grow 
rapidly and exceed the background around $\sqrt{\hat{s}} = (700 \sim 800)\gev$
which is less than the energy cutoff $\varLambda_i = 1\tev$.

The mono-$\gamma$ process has been extensively studied at
LHC for various DS particle models containing a mediator \cite{ATLAS:2020uiq}.
It is expected that the same data can give exclusion limits on our model at the
same level, \ie, $\varLambda_i \sim m_{\rm med} $. 
Our simulations are conducted using the toolboxes \textsf{MadGraph} \cite{Alwall:2014hca,Frederix:2018nkq}
and \textsf{FeynRules} \cite{Alloul:2013bka}.
The signal total cross sections are estimated according to the 
event selections given in the ATLAS search \cite{ATLAS:2020uiq}.
A photon with a transverse energy above 130\,GeV was required 
at the matrix-element level.
A strong kinematical cut $\met \gtrsim 200\gev$ is used to select the signal events. 
In this region, the parton shower effect is negligible.
So our simulation is done at the generator level.
The total detection efficiency is accounted for by an overall normalization factor
which is estimated by validating the irreducible background process 
$pp\to\gamma Z(\nu\nu)$. Consequently, the detector-level predictions 
are obtained by multiplying this normalization factor with the 
generator-level results.

\begin{figure}[t]
\centering
\includegraphics[width=0.44\textwidth]{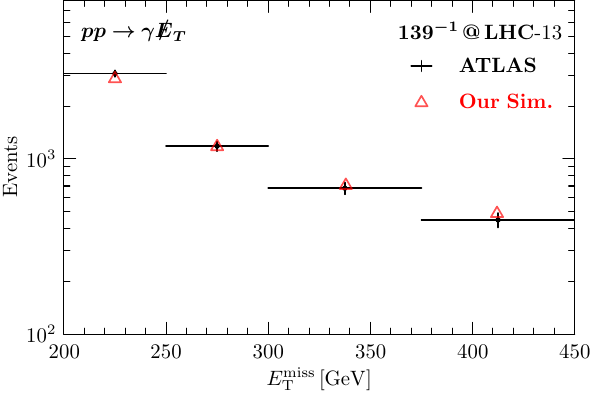}
\includegraphics[width=0.44\textwidth]{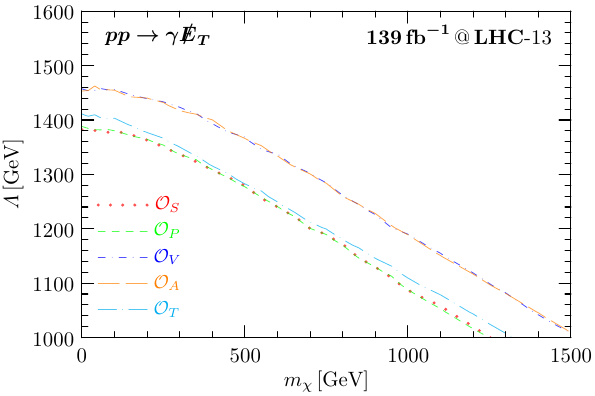}
\caption{\it 
\textbf{Light panel}: Validation of our simulation for the missing transverse momentum
distribution. The experimental data are (black line) taken from Ref. \cite{ATLAS:2020uiq}
and our results (red triangle) have been renormalized by multiplying an overall constant.
\textbf{Right panel}: The expected 95\%\,C.L. exclusion limits at LHC-13.
}
\label{fig:valid:ma:LHC13}
\end{figure}
The left panel of \gfig{fig:valid:ma:LHC13} compares
the missing energy distributions between our results and the ATLAS data
for the total background (sum of the irreducible and reducible contributions). 
Our prediction on the total background is obtained  
by rescaling the irreducible background by an overall renormalization factor
to match with the total background event number. 
One can see an excellent match in the missing energy distribution 
and it is clear that within the experimental uncertainty our simulation is valid.
The above results indicate that the approximation of an overall normalization factor 
works well for both the total event number and the differential distributions. 
This excellent approximation is also employed to estimate the expected exclusion limits 
on the four-fermion effective operators.

The right panel of \gfig{fig:valid:ma:LHC13} shows the expected exclusion limits 
at 95\%\,C.L. in the $m_\chi - \varLambda$ plane. We can see that the strongest limit is
given for the (axial)-vector operator, the constraint on the tensor operator is slightly weaker,
and the weakest one is for the (pseudo)-scalar operator. However, the differences are not very 
significant. In case of $m_\chi \sim 0$, the lower bounds are roughly $1.4\tev$.
On the other hand, for $\varLambda_i = 1\tev$, a heavy dark fermion with mass
from $1.2\tev$ to $1.5\tev$ can be excluded.

\subsection{Mono-Jet}
\label{sec:monoj}
\begin{figure}[t]
\centering
\includegraphics[width=0.3\textwidth]{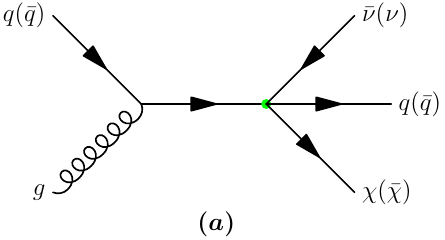}
\quad
\includegraphics[width=0.3\textwidth]{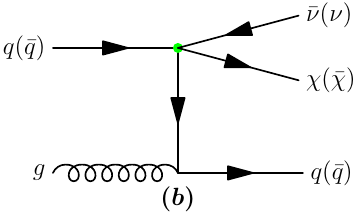}
\quad
\includegraphics[width=0.3\textwidth]{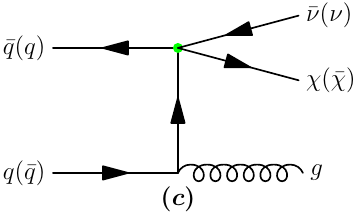}
\\[2mm]
\includegraphics[width=0.3\textwidth]{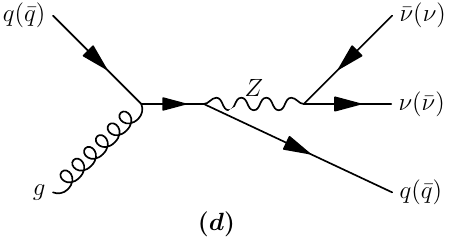}
\quad
\includegraphics[width=0.3\textwidth]{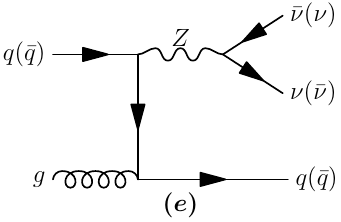}
\quad
\includegraphics[width=0.3\textwidth]{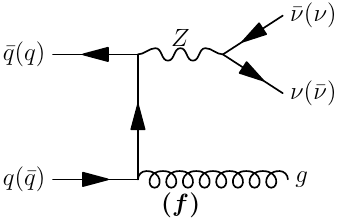}
\caption{\it 
Feynman diagrams contributing to the mono-jet events. The panels
\textbf{(a)}, \textbf{(b)} and \textbf{(c)} are for the signal operators
(green dot) while
\textbf{(d)}, \textbf{(e)} and \textbf{(f)} are for the irreducible background.
}
\label{fig:feyn:monoj:LHC}
\end{figure}

The effective operators in \geqn{eq:effo} can also induce
mono-jet events. There are three channels that can contribute as signals
and the corresponding Feynman diagrams are shown in \gfig{fig:feyn:monoj:LHC}\,(a), (b) and (c).
In contrast to the mono-$\gamma$ process, the mono-jet channel can also be initiated 
by the gluon components of the incoming hadrons. Hence, stronger constraints are expected. From \gfig{fig:feyn:monoj:LHC}, one can see that 
except for the $s$-channel (as shown in the panel (a)), all the other channels are
initiated by the initial-state radiation (as shown in the panels (b) and (c)). 
Hence it is expected that the signals are
populated at the forward and backward regions.

This is also true for the major background. The
associated production of a jet and a $Z$ boson followed 
by the invisible decays $Z\to\nu\bar\nu$ provides the
irreducible background whose Feynman diagrams
are shown in the \gfig{fig:feyn:monoj:LHC}\,(d), (e) and (f)
panels.
There are also reducible channels, for instance $q\bar{q} \to W(\tau\nu)$
with the subsequent leptonic decay into $\tau$ that can
also contribute to the total background events.
In case of $\sqrt{s}=13\tev$, the total reducible backgrounds contribute
about $40\%$ of the total mono-jet events \cite{ATLAS:2021kxv}.
Here we discuss the physical properties of
only the irreducible background, 
\ie, the process $q\bar q'/qg \to j Z(\nu\bar\nu)$.
In this case, the $Z$ boson comes from either the initial- or
final-state radiation. Hence the irreducible backgrounds are also
expected to be populated at the forward and backward regions.

\begin{figure}[t]
\centering
\includegraphics[width=0.32\textwidth]{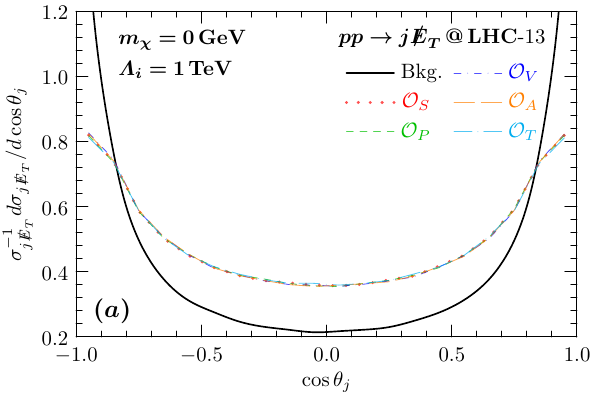}
\hfill
\includegraphics[width=0.32\textwidth]{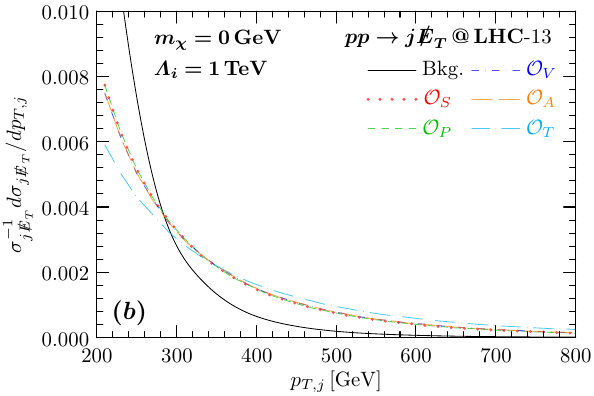}
\hfill
\includegraphics[width=0.32\textwidth]{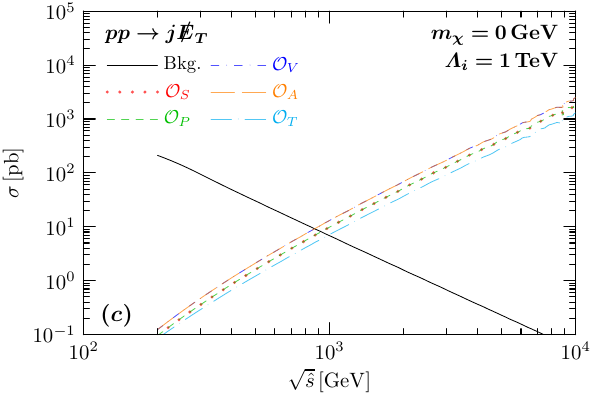}
\caption{\it 
The normalized parton-level distributions of the polar angle ($\theta_j$) \textbf{(a)}
and transverse momentum ($p_{T,j}$) \textbf{(b)} 
in the laboratory frame with center of mass energy $\sqrt{s} = 13\tev$. 
\textbf{(c):} The signal and background total cross sections as functions of the
center-of-mass energy at the parton level, $\sqrt{\hat{s}}$. 
In all three panels, the signal (colorful non-solid curves) are shown for parameters 
$m_\chi = 0\gev$ and $\varLambda_i = 1\tev$, 
and the background (black-solid curve) stands for the irreducible 
channel $q\bar q'/qg \to j Z(\nu\nu)$.
}
\label{fig:ca:pt:mj:LHC13}
\end{figure}

The \gfig{fig:ca:pt:mj:LHC13}\,(a) and (b) show the normalized
jet polar angle 
($\theta_j$) and transverse momentum ($p_{T,j}$) distributions 
in the laboratory frame, respectively. The signal properties
(colorful non-solid curves) are shown 
for parameters $m_\chi = 0\gev$ and $\varLambda_i = 1\tev$,
and the irreducible background is shown as black-solid curve.
Both signals and background are dominant at
the forward and backward regions, while the background has
a more ``collinear singularity" behavior. On the other hand,
the signal operators having different Lorentz structure possess 
completely the same polar angle distribution. 
So the polar angle cannot distinguish the signal operators.
The \gfig{fig:ca:pt:mj:LHC13}\,(b) panel shows that 
the signal events are dominant at the soft $p_{T,j}$ region 
and there are some differences 
among the five signal operators. Particularly, at the large $p_{T,j}$ region, 
the distribution for the tensor operator is slightly larger
than those for other operators.
It turns out that the mono-jet search is more sensitive to the tensor operator.
Furthermore, the signal distributions in the
region $p_{T,j} \gtrsim 300\gev$ are larger than the background one.
Hence it is a good cut to enhance the signal significance. 
\gfig{fig:ca:pt:mj:LHC13}\,(c) shows the variation of
the total parton-level cross sections with respect to the 
center-of-mass energy $\sqrt{\hat{s}}$ for both signal and 
the irreducible background.
While the background decreases quickly with increasing 
center-of-mass energy $\sqrt{\hat{s}}$, the signal total
cross sections grow rapidly. Similar to the mono-$\gamma$
process, signals exceed the background around $\sqrt{\hat{s}} = 900\gev$.

The ATLAS collaboration searched for new phenomena events containing 
an energetic jet and a large missing transverse momentum \cite{ATLAS:2021kxv}.
For an axial-vector mediated model, the exclusion limit for $m_\chi =0 \gev$ 
reaches about $2.1\tev$. It is expected that the four-fermion contact couplings
can be constrained at a similar level. We use those data to
constrain the parameters $m_\chi$ and $\varLambda_i$.
In our analysis, events are selected according to the signal
region definitions,
$E_{T}^{\rm miss} > 200\gev$, $p_{T, j} > 150$\,GeV
and $|\eta_j| < 2.4$, in \cite{ATLAS:2021kxv}. 
With strong cut on the missing transverse energy,
the parton shower effect can be ignored and hence
our simulation is done at the generator level.
The total detection efficiency is taken into account by an overall 
normalization factor which is estimated by validating the irreducible 
background process $pp\to j Z(\nu\nu)$. This overall normalization factor
approximation approach is employed to estimate the detector-level prediction 
for both signal and background.

\begin{figure}[t]
\centering
\includegraphics[width=0.44\textwidth]{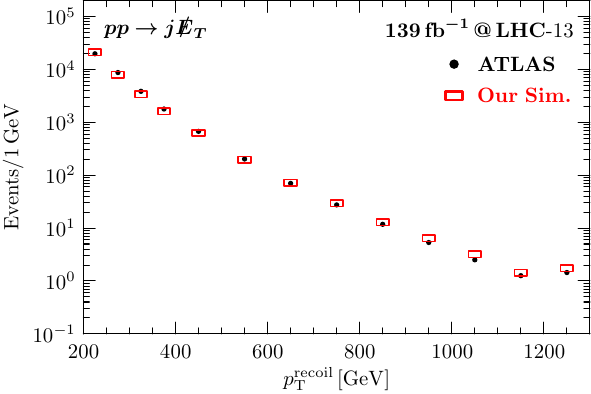}
\includegraphics[width=0.44\textwidth]{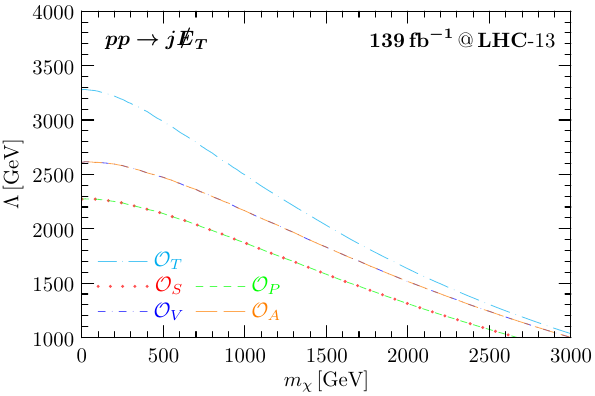}
\caption{\it 
\textbf{Left panel}: Validation of our simulation for the $p_{T}^{\rm recoil}$ 
distribution of the irreducible background
channel $pp\to j Z(\nu\bar\nu)$ at LHC with a total luminosity $\call = 139\fb^{-1}$.
The experimental data (black dots) are taken from \cite{ATLAS:2021kxv}
and our results (red rectangles) have been renormalized by multiplying an overall constant.
\textbf{Right panel}: The expected 95\%\,C.L. exclusion limits at LHC-13.
}
\label{fig:valid:mj:LHC13}
\end{figure}

The left panel of \gfig{fig:valid:mj:LHC13} shows the comparison between 
our simulation (red square) and the ATLAS result (black dot)
for the $p_{T}^{\rm recoil}$ (which is equivelent to the jet
transverse missing energy or transverse momentum at the generator level)
distribution of the irreducible background
channel $pp\to j Z(\nu\bar\nu)$ at LHC-13 with a total luminosity
of $\call = 139\fb^{-1}$.
The detector effects can be well described 
in the interested $p_{T}^{\rm recoil}$ region by an overall 
normalization factor. Within the experimental uncertainty, 
our simulation is valid. The same normalization factor will be multiplied 
to the signal cross section to estimate the exclusion limits.
It is clear that the approximation of an overall 
normalization factor works well not only for the total event number 
but also the differential distributions. This is particularly important 
when estimating the exclusion limit with 
$\chi^2$ for the $p_{T}^{\rm recoil}$ distribution.

The right panel of \gfig{fig:valid:mj:LHC13} shows the 95\% expected 
exclusion limits in the $m_\chi$-$\varLambda$ plane for our signal operators.
We can see that the strongest limit is given for the tensor operator, 
which can reach about $3.3\tev$ for $m_\chi \sim 0$. 
This is because of larger cross section (compared to the other operators) 
and also more events are populated at the large $p_{T,j}$ region, 
as shown in \gfig{fig:ca:pt:mj:LHC13}\,(b).
On the other hand, for $\varLambda_i = 1\tev$, a heavy dark fermion 
with mass from $2.6\tev$ to $3\tev$ can be excluded. 
The constraint on the (axial)-vector operator is slightly weaker, 
$2.6\tev$ for $m_\chi \sim 0$. 
The weakest one is for the (pseudo)-scalar operator, 
roughly $2.3\tev$ for $m_\chi \sim 0$. One can also see that, 
the exclusion limits are about $2$ times stronger than the
mono-$\gamma$ ones.
This is mainly due to the considerably large production cross section
of the mono-$j$ process than its mono-$\gamma$ counterpart,
as can be seen by comparing \gfig{fig:ca:pt:mj:LHC13}\,(c) and
\gfig{fig:ca:pt:ma:LHC13}\,(c).

\subsection{Mono-$Z$}
\label{sec:monoz}

\begin{figure}[ht]
\centering
\includegraphics[width=0.3\textwidth]{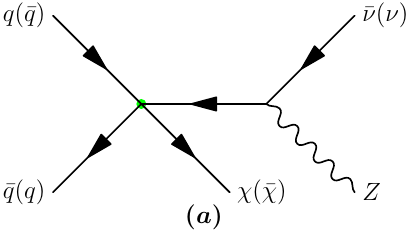}
\quad
\includegraphics[width=0.3\textwidth]{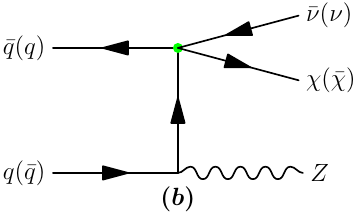}
\\
\includegraphics[width=0.3\textwidth]{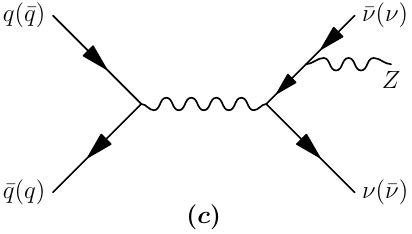}
\quad
\includegraphics[width=0.3\textwidth]{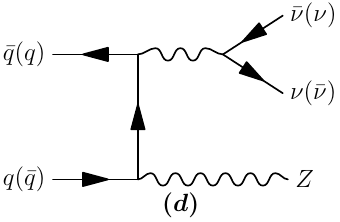}
\caption{\it 
Feynman diagrams for the mono-$Z$ events,
\textbf{(a)} and \textbf{(b)} for the signal while 
\textbf{(c)} and \textbf{(d)} for the irreducible background.
}
\label{fig:feyn:mz:LHC}
\end{figure}

The effective operators in \geqn{eq:effo} can also induce
mono-$Z$ events. There are two channels that can contribute to 
the signals as shown in 
\gfig{fig:feyn:mz:LHC}\,(a) and (b). Similarly, there are also two channels 
for the major irreducible background $pp \to Z \nu\bar\nu$, 
as depicted in \gfig{fig:feyn:mz:LHC}\,(c) and (d).
In both cases, the $Z$ boson is emitted either by the incoming quark or 
by the outgoing neutrino. Hence it is expected that both the signal and 
background are populated in the forward and backward regions. 

\begin{figure}[b]
\centering
\includegraphics[width=0.32\textwidth]{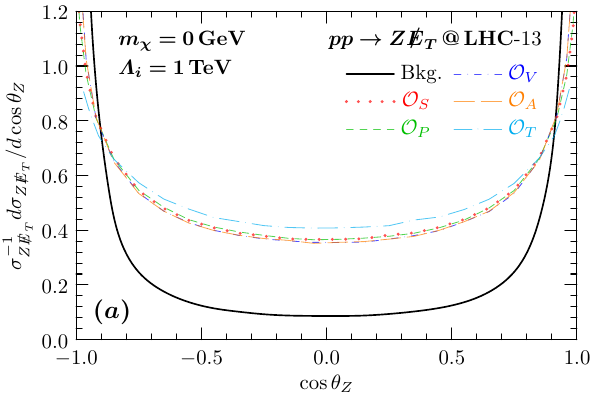}
\hfill
\includegraphics[width=0.32\textwidth]{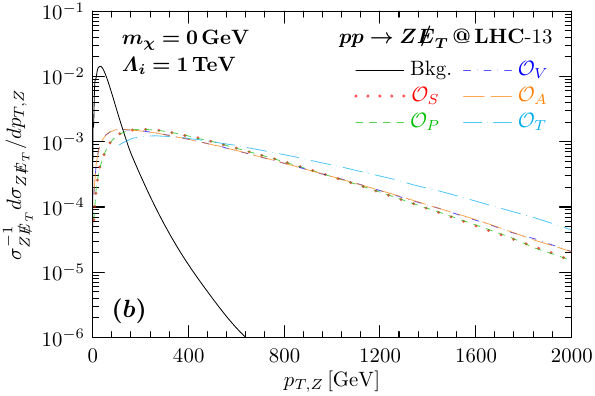}
\hfill
\includegraphics[width=0.32\textwidth]{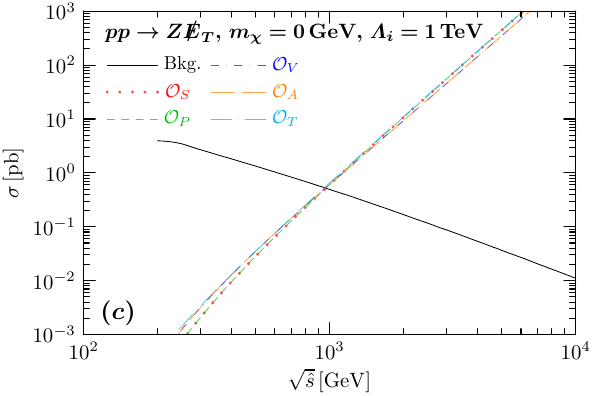}
\caption{\it 
The normalized parton-level distributions of the polar angle ($\theta_Z$, \textbf{panel (a)})
and transverse momentum ($p_{T,Z}$, \textbf{panel (b)}) of the $Z$ boson
in the laboratory frame with center-of-mass energy $\sqrt{s} = 13\tev$. 
\textbf{Panel (c):} the parton-level background and signal total cross sections as functions of the
center-of-mass energy $\sqrt{\hat{s}}$. 
The signal (colorful non-solid curves) are shown for parameters 
$m_\chi = 0\gev$ and $\varLambda_i = 1\tev$, 
and the background (black-solid curve) stands for the irreducible
contribution from the channel $q\bar q\to Z\nu\nu$.
}
\label{fig:ca:pt:mz:inc:LHC13}
\end{figure}
The \gfig{fig:ca:pt:mz:inc:LHC13}\,(a) and (b) panels
show the normalized $Z$ polar angle 
($\theta_Z$) and transverse momentum ($p_{T, Z}$) distributions
in the laboratory frame, respectively. The signals with model parameters 
$m_\chi = 0\gev$ and $\varLambda_i=1\tev$ are shown as colorful non-solid curves while
the irreducible background contribution 
$pp \to Z \nu\bar\nu$ as black-solid curve.
As expected, both signals and the irreducible background are dominant 
in the forward and backward regions. For comparison,
the background has a more collinear behavior. 
The reason is twofold. Firstly, the dominant contribution to the irreducible background
comes from the $t$-channel di-boson production $q\bar q\to ZZ(\nu\nu)$,
as shown in \gfig{fig:feyn:mz:LHC}\,(d).
Secondly, compared to the signals, the $s$-channel contribution of the background 
has a stronger suppression in the large $\hat{s}$ region.
One can also see that there are some differences in the polar angle distribution
among the signal operators with different Lorentz structures, 
particularly in the forward and backward regions. 
However, this difference can be dismissed when distinguishing
the type of the signal operators 
because of the large background in this region.
Furthermore, as one can see from
\gfig{fig:ca:pt:mz:inc:LHC13}\,(b) that 
the signal events are dominant in the large transverse momentum region
and the signals exceed the background in the region $p_{T,Z} \gtrsim 200\gev$.
So the dominant contribution to the signal significance 
comes from the events having relatively large transverse momentum.
\gfig{fig:ca:pt:mz:inc:LHC13}\,(c) shows how
the total parton-level cross sections depend on the center-of-mass energy 
$\sqrt{\hat{s}}$ for both signals and the irreducible background.
One can clearly see that while the background decreases quickly with the increasing 
center-of-mass energy $\sqrt{\hat{s}}$, the signal total cross sections grow 
rapidly with $\sqrt{\hat{s}}$. 
Again, signals exceed the background around $\sqrt{\hat{s}} = 1\tev$.

However, it is not straightforward to study the experimental constraints on the 
model parameters, because the $Z$ reconstruction is necessary in practical 
measurement. For its leptonic decay modes, since tracks can be measured precisely,
the mono-$Z$ events can be efficiently selected by putting a cut on the lepton pair invariant mass. However, it is not true for the hadronic decay modes. On one hand,
the jet momentum uncertainty is much larger than its lepton counterpart.
On the other hand, both the electroweak (for instance $pp\to W^{\pm}(jj) + \met$)
and the pure QCD channels $pp\to jj +\met$ can contribute as background.
Actually, the pure QCD contribution completely dominates
the total background \cite{ATLAS:2018nda}. So we study the
leptonic and hadronic decay modes separately in \gsec{sec:monoz:lep}
and \gsec{sec:monoz:had}.

\subsubsection{Leptonic decay modes}
\label{sec:monoz:lep}

Let us first study the mono-$Z$ production followed by
leptonic decays $Z\to\ell\ell$ ($\ell = e$ and $\mu$) 
at the center-of-mass energy $\sqrt{s}=13\tev$. 
\begin{figure}[ht]
\centering
\includegraphics[width=0.44\textwidth]{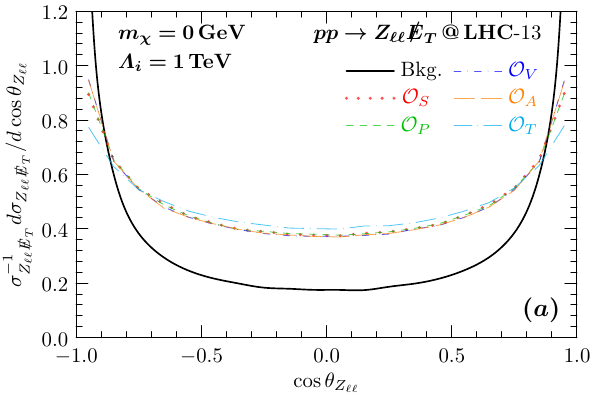}
\quad
\includegraphics[width=0.44\textwidth]{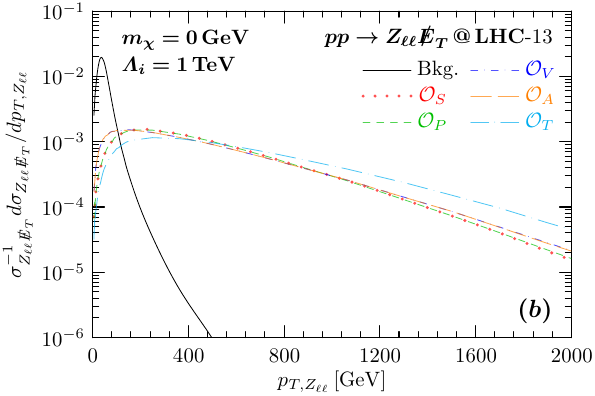}
\caption{\it 
The normalized distributions of the polar angle ($\theta_{Z_\ell}$, \textbf{left panel}) and
the transverse momentum ($p_{T,Z}$, \textbf{right panel}) of the reconstructed 
$Z$ boson from its leptonic decay modes. 
In both panels, the observables are calculated in the laboratory frame
with the center-of-mass energy 
$\sqrt{s} = 13\tev$. The signals (colorful non-solid curves) are shown for parameters 
$m_\chi = 0\gev$ and $\varLambda_i = 1\tev$ while the background (black-solid curve) 
stands for the irreducible contribution from $q\bar q\to Z\nu\nu$.
}
\label{fig:ca:pt:mz:lep:LHC13}
\end{figure}
The left and right panels of \gfig{fig:ca:pt:mz:lep:LHC13} show
the normalized polar angle and transverse momentum distributions 
of the reconstructed $Z$-boson from the two leptons. 
One can clearly see that for both the signal (colorful non-solid curves) 
and the irreducible background (black-solid curve),
the distinctive kinematic properties (that has been discussed
for the on-shell $Z$ production) 
of the mono-$Z$ events can be readily reconstructed.

The ATLAS collaboration has searched for dark matter 
in the mono-$Z$ production with the $Z$ boson decaying to two leptons 
\cite{ATLAS:2021gcn}. The major backgrounds come from the irreducible 
contribution of the resonant $pp\to ZZ \to \ell\ell\nu\nu$ and
non-resonant ($pp\to WW \to \ell\ell\nu\nu$, \etc) channels.
In addition, the reducible backgrounds come from the channels
$pp \to WZ$ and $pp \to Z + {\rm jets}$.
The events are selected by requiring that the leptons have 
$p_{T} > 20, 30$\,GeV when ordered with increasing $p_T$
and their distance has to fulfill $\Delta R_{\ell\ell} < 1.8$. 
Furthermore, only those events with
$E_{T}^{miss} > 90\gev$ and containing exactly two oppositely charged 
electrons or muons with an invariant mass
$76\gev < m_{\ell\ell} < 106\gev$ around the $Z$ boson
mass are selected for further 
analysis. The signal region is defined by a selection condition 
$m_T \ge 200\gev$ on the transverse mass $m_T$,
\bee
m_T
\equiv
\sqrt{
\left[ \sqrt{m_Z^2 + (p_{T}^{\ell\ell})^2 } + \sqrt{m_Z^2 + (E_{T}^{\rm miss})^2 }\right]^{2}
-\left[ \vec{p}_{T}^{\;\ell\ell} + \vec{E}_{T}^{\rm miss} \right]^2
}\,,
\ene
to select signals from the backgrounds. Since at the parton level, 
$ \vec{p}_{T}^{\;\ell\ell} = - \vec{E}_{T}^{\rm miss}$, 
hence the transverse mass is simply given as
$m_T = 2 \sqrt{m_Z^2 + (p_{T}^{\ell\ell})^2}$. 
As a result, the requirement $m_T \ge 200\gev$ is equivalent to
$p_{T}^{\ell\ell} \gtrsim 85\gev$ which is a relatively strong
cut for the parton shower effect.
Hence it is almost equivalent to do the simulation at the generator level since the parton shower effect would not be important with cut.

\begin{figure}[t]
\centering
\includegraphics[width=0.44\textwidth]{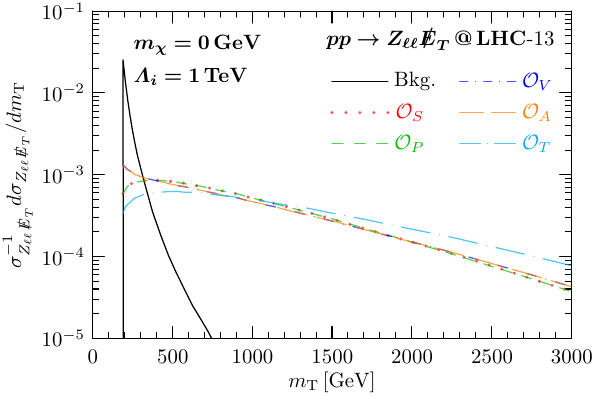}
\quad
\includegraphics[width=0.44\textwidth]{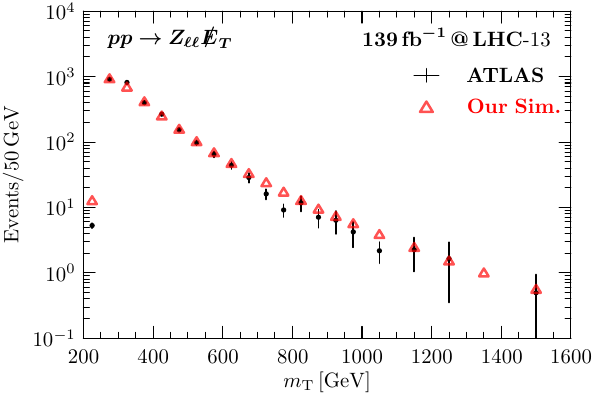}
\caption{\it 
\textbf{Left panel:} The normalized transverse mass $m_T$
distribution using the $Z$ leptonic decay modes
($Z\to\ell\ell$ with $\ell = e,\mu$). The signal distributions
(colorful non-solid curves) are illustrated with 
$m_\chi = 0\gev$ and $\varLambda_i = 1\tev$ while the
background (black-solid curve) stands for the irreducible
contribution from $pp\to Z(\ell\ell)\nu\bar\nu$.
\textbf{Right panel:} The $m_T$ distribution of the irreducible
background at LHC-13. The experimental data are (black dot)
taken from the Ref.\,\cite{ATLAS:2021gcn} and our results
(red trangle) have been renormalized by multiplying an
overall normalization factor.
}
\label{fig:mt:mz:lep:LHC}
\end{figure}
The left panel of \gfig{fig:mt:mz:lep:LHC} shows the
normalized transverse mass $m_T$ distributions for the signals 
(colorful non-solid curves) with $m_\chi = 0\gev$ and 
$\varLambda_i = 1\tev$ as well as the irreducible background
(black-solid curve) 
from $pp\to Z(\ell\ell) \nu\nu$. One can clearly see that 
the irreducible background drops very quickly with 
increasing $m_T$ but there are sizable long tails for signals.
Hence $m_T$ is really a good observable 
for selecting the signal events. The total detector efficiency is approximated
by an overall normalization factor that is estimated  
by validating the irreducible background $pp\to Z(\ell\ell) \nu\nu$.
The right panel of the \gfig{fig:mt:mz:lep:LHC} compares 
our simulation (red trangle) with the ATLAS data (black dot) 
for the $m_T$ distribution of the irreducible background
$pp\to Z(\ell\ell) \nu\nu$ at LHC-13.
One can see that the detector effects can be well modeled (for both
the total event number and the differential distibution) 
by an overall normalization factor. Again we assume that
the total detector
efficiency is universal for both signals and background. Namely,
both the total event number and the signal $m_T$ distribution
at the detector level are obtained by multiplying the
corresponding values at the generator level with the same
normalization factor.

\begin{figure}[bht]
\centering
\includegraphics[width=0.48\textwidth]{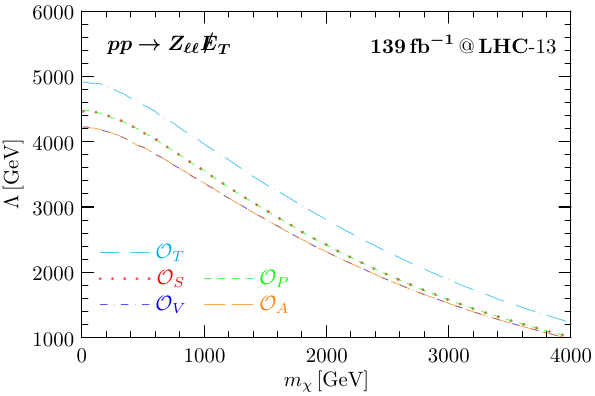}
\caption{The expected exclusion limits at 95\%\,C.L. using
the mono-$Z$ events at LHC-13. The $Z$-boson is reconstructed
by its leptonic decay modes $Z\to\ell\ell$ with $\ell = e,\mu$.}
\label{fig:exlimit95:mz:lep:LHC13}
\end{figure}

\gfig{fig:exlimit95:mz:lep:LHC13} shows the 95\%\,C.L.
expected exclusion limits on the $m_\chi$-$\varLambda$ plane.
The strongest limit is given for the tensor operator and 
it can reach about $4.9\tev$ for $m_\chi \sim 0$. This
is because of the relatively larger cross section than the other
operators (as shown in \gfig{fig:ca:pt:mz:inc:LHC13} (c))
and also more events are populated in the large $m_T$ region 
(as shown in the left panel of \gfig{fig:mt:mz:lep:LHC}). 
The constraint on the (pseudo)-scalar operator 
is slightly weaker which is about $4.5\tev$ for $m_\chi \sim 0$. 
The weakest one is for the (axial)-vector operator,
being roughly $4.2\tev$ 
for $m_\chi \sim 0$. On the other hand, for $\varLambda_i = 1\tev$, 
a heavy dark fermion with mass up to $4\tev$ can be excluded. 
One can also see that the exclusion limits are stronger than
the mono-$j$ (and the mono-$\gamma$) process, even through
the mono-$j$ process has considerably larger cross section which
can be seen by comparing \gfig{fig:ca:pt:mz:inc:LHC13}\,(c)
with \gfig{fig:ca:pt:mj:LHC13}\,(c) and
\gfig{fig:ca:pt:ma:LHC13}\,(c). This is mainly due to the
significantly larger background and uncertainty of the
mono-$j$ event.
Comparing with the mono-$\gamma$ process, the total cross sections are
at the same level but the $Z$ transverse momentum distribution 
(and hence missing transverse momentum and transverse mass) 
is sizably harder than the photon one.
Therefore, we have larger signal significance in the mono-$Z_{\ell\ell}$ event
and hence stronger exclusion limits.

\subsubsection{Hadronic decay modes}
\label{sec:monoz:had}

\begin{figure}[hb]
\centering
\includegraphics[width=0.44\textwidth]{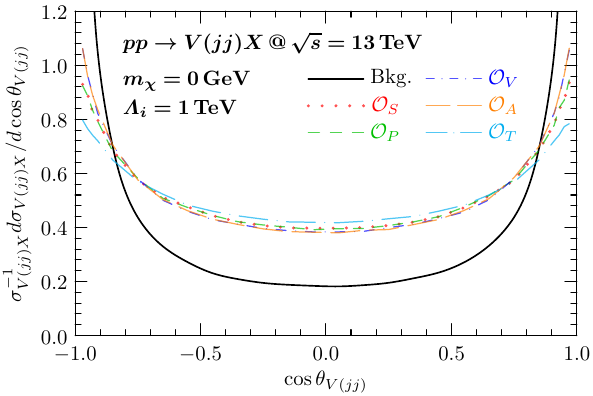}
\quad
\includegraphics[width=0.44\textwidth]{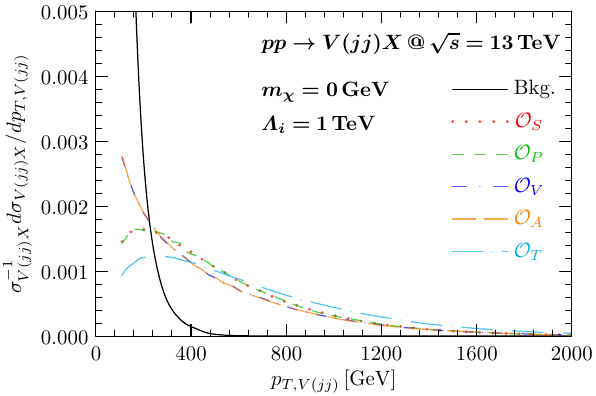}
\caption{\it 
The normalized polar angle (\textbf{left panel}) and transverse momentum 
(\textbf{right panel}) distributions of the reconstructed vector boson 
from two jets
with invariant mass $m_{jj} \in [65,\, 105]\gev$
for two channels $pp\to V(jj)+\met$ ($V=Z$ and $W$) at 
center-of-mass energy $\sqrt{s} = 13\tev$. 
In both panels, the signal events with parameters $m_\chi = 0\gev$ and 
$\varLambda_i=1\tev$ are illustrated by colorful non-solid curves
and the background is shown as black-solid curve.
}
\label{fig:ca:pt:mz:vjj:LHC13}
\end{figure}

The mono-$Z$ event reconstruction for the hadronic decay mode 
is much more complex than the leptonic one.
Firstly, because of the relatively small mass difference between the $W$ 
and $Z$ bosons as well as the large jet momentum uncertainty, 
the reconstructed $Z$ bosons are inevitably contaminated 
by the hadronic $W$ decay products.
\gfig{fig:ca:pt:mz:vjj:LHC13} shows the normalized polar angle
$\cos \theta_{V(jj)}$ and transverse momentum $p_{T, V(jj)}$
distributions of the reconstructed vector boson 
from $pp\to V(jj)+\met$ for both $V=Z$ and $W$. Comparing with the purely 
mono-$Z$ contribution in \gfig{fig:ca:pt:mz:inc:LHC13}\,(a),
the background is less towards the forward and backward regions.
This is due to the large jet transverse energy cut,
as shown in \gfig{fig:ca:pt:mz:vjj:LHC13}\,(b).
\begin{figure}[h]
\centering
\includegraphics[width=0.44\textwidth]{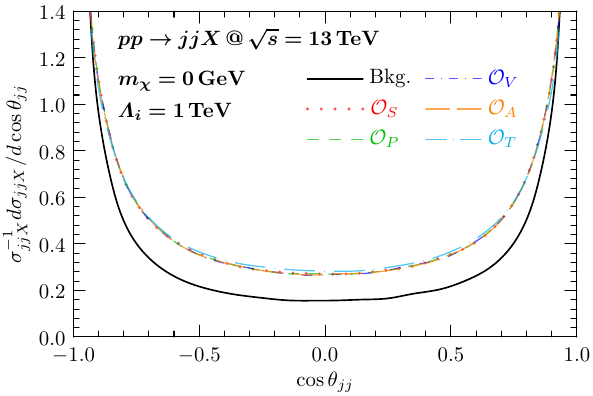}
\quad
\includegraphics[width=0.44\textwidth]{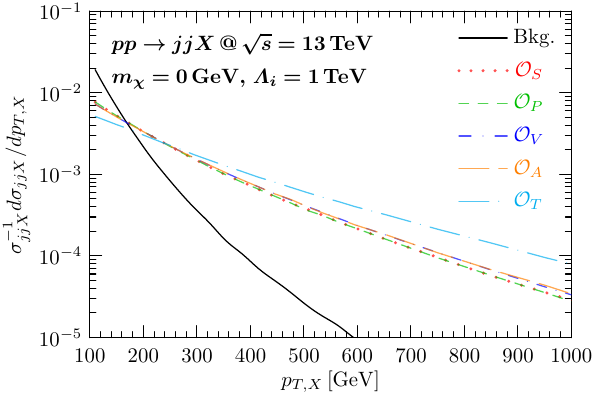}
\caption{\it 
The normalized polar angle (left) and transverse momentum (right) distributions 
in the invariant mass window $m_{\rm jets} \in [65,\, 105]\gev$ for the channels 
$pp \to \met + {\rm jets}$ with QCD jets, respectively. For all the panels, 
the signal events (colorful non-solid curves) are illustrated with parameters  
$m_\chi = 0\gev$ and $\varLambda_i=1\tev$.
}
\label{fig:ca:pt:mz:jj:LHC13}
\end{figure}
Secondly, the mono-$Z$ events are even more heavily contaminated 
by the associated production of missing energy with pure QCD jets.
The dominant contribution comes from the associated production of 2-jets
with an invisibly decaying $Z$ boson, \ie, 
the channels $pp \to Z(\nu\nu) + jj$. 
In this case, the 2 jets with invariant mass in the window 
$m_{\rm jets} \in [65,\, 105]\gev$ can be miss-identified as a hadronically 
decaying $Z$ boson. Actually, it consists majority of the total
background \cite{ATLAS:2018nda}. 
\gfig{fig:ca:pt:mz:jj:LHC13}
shows the polar angle and transverse momentum distributions
of the summed jet ($jj$) for both the signal processes 
$pp \to \chi\nu + 2j$ and the dominant background
$pp \to Z(\nu\nu) + 2j$. Comparing with \gfig{fig:ca:pt:mz:vjj:LHC13},
both signal and background are pushed towards the forward
and backward regions. Consequently, the transverse momentum
of the fake $Z$-boson from QCD jets are softer.

\begin{figure}[bh]
\centering
\includegraphics[width=0.44\textwidth]{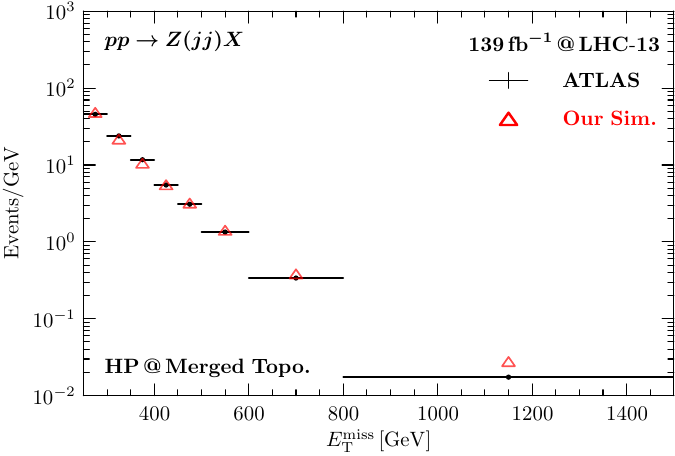}
\quad
\includegraphics[width=0.44\textwidth]{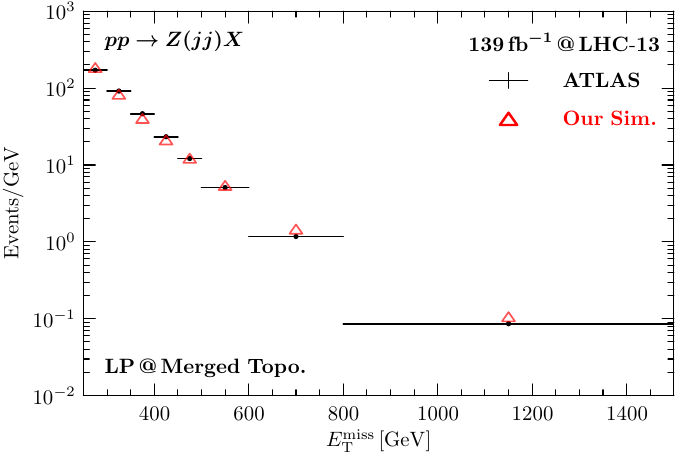}
\caption{\it 
Validation of our simulation for the missing transverse energy distribution of the 
pure QCD di-jet production channel $pp\to Z(\nu\nu) + jj$ at the LHC-13
with a total luminosity $\call = 36.1\fb^{-1}$. The \textbf{left} and \textbf{right} panels
are for the HP and LP regions, respectively. 
The experimental data \cite{ATLAS:2018nda} (black-solid line)
and our results (red triangle) have been normalized 
by multiplying an overall constant.
}
\label{fig:valid:mz:had:LHC}
\end{figure}

The ATLAS collaboration has searched for dark matter in the mono-$Z/W$ production
with hadronically decaying $Z/W$-boson~\cite{ATLAS:2018nda} in three  categorizations:
the high purity (HP) and low purity (LP) regions for merged topology (MT), 
and the resolved topology (RT). In each region, 3 configurations are
defined according to $b$-tagging. Since the background in the RT region
is about 1 order larger, our following calculations consider
only the LP and HP regions. Furthermore, the contributions with non-zero
$b$-tagging number are sub-leading and hence can be ignored.
\gfig{fig:valid:mz:had:LHC} compares 
our simulation (red triangle) and the ATLAS result (black-solid line) 
for the missing transverse energy distribution of the pure QCD di-jet
production channel $pp\to Z(\nu\nu) + jj$ at the LHC-13 with a total
luminosity $\call = 36.1\fb^{-1}$.
The total event number of our result has been normalized to the experimental data.
One can see that the detector effects can be properly described by an overall 
normalization factor. The same normalization factor is then multiplied with the signal 
cross sections to calculate the corresponding exclusion limits.
\begin{figure}[h]
\centering
\includegraphics[width=0.44\textwidth]{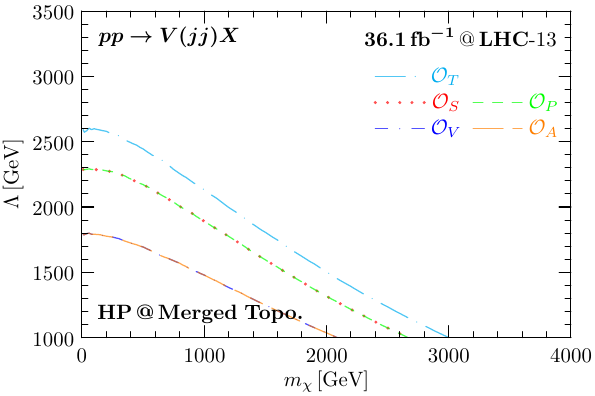}
\;
\includegraphics[width=0.44\textwidth]{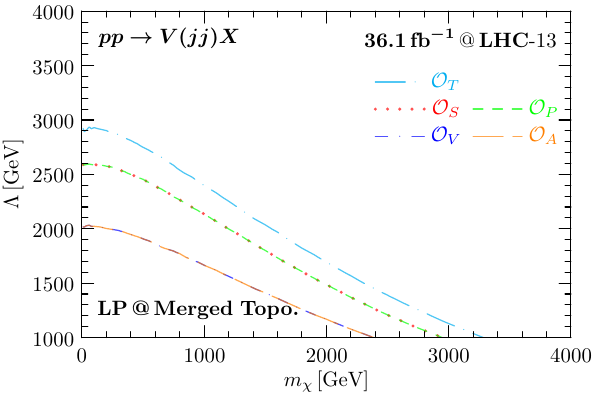}
\caption{\it 
The expected exclusion limits at 95\%\,C.L. using the mono-$Z/W$ events with hadronically
decaying $Z/W$ boson at LHC-13. The \textbf{left} and \textbf{right}
panels are for the HP and LP regions, respectively. 
}
\label{fig:sens:mz:had:LHC}
\end{figure}
\gfig{fig:sens:mz:had:LHC} shows the expected 
exclusion limits at 95\%\,C.L. in the $m_\chi$-$\varLambda$ plane.
The strongest limit comes from the tensor operator. With 
$m_\chi \sim 0$, the lower limits can reach $2.6$ and $2.9\,\tev$ in 
the HP and LP regions, respectively. For comparison, the
constraint on the (pseudo)-scalar operator can reach
$2.3$ and $2.6\tev$ in the HP and LP regions, respectively,
with $m_\chi \sim 0$ is slightly weaker.
The weakest constraint happens for the (axial)-vector operator
which is only about $1.8$ and $2.0\tev$ in the HP and LP regions,
respectively. 
On the other hand, for the energy scale $\varLambda_i = 1\tev$, 
a heavy dark fermion with mass from $2\tev$ to $4\tev$ can be excluded.

\subsection{Projected Sensitivities at Future Upgrades}
\label{sec:future}
This section studies the projected sensitivities at the upgraded
versions of LHC \cite{CidVidal:2018eel}.
As discussed in the last sections, the signal cross sections grow with
the center-of-mass energy while the background one decreases.
It is then expected that the future upgrades of LHC have
great advantages for probing the four-fermion absorption operators.
Furthermore, the upgrades will accumulate much larger luminosity.
\gtab{tab:hptc} lists the studied configurations and
the parton-level kinematic cuts.
\begin{table}[h]
\renewcommand\arraystretch{1.35}
\begin{center}
\begin{tabular}{c|c c c}
{\rm Process}   & 14~TeV, 3\iab (LHC-14) & 25~TeV, 20\iab (LHC-25)
\\\hline\hline
$pp\to \gamma\met$ 
& 
\begin{tabular}{c} 
$p_{T,\gamma} \geqslant 200\,\gev$\,,\;
$\big|\eta_{\gamma}\big| \leqslant 2.5$
\\ 
$ \met \geqslant 200\,\gev$ 
\end{tabular}
& 
\begin{tabular}{c} 
$p_{T,\gamma} \geqslant 400\,\gev$\,,\;
$\big|\eta_{\gamma}\big| \leqslant 2.5$
\\ 
$ \met \geqslant 400\,\gev$ 
\end{tabular}
\\[2mm]\hline
$pp\to j\met$ 
& 
\begin{tabular}{c} 
$p_{T,j} \geqslant 200\,\gev$\,,\; $\big|\eta_{j}\big| \leqslant 2.4$ 
\\ 
$ \met \geqslant 200 \, \gev$ 
\end{tabular}
& 
\begin{tabular}{c} 
$p_{T,j} \geqslant 400\,\gev$\,,\; $\big|\eta_{j}\big| \leqslant 2.4$ 
\\ 
$ \met \geqslant 400 \, \gev$ 
\end{tabular}
\\[2mm]\hline
$pp\to Z(\ell^{+}\ell^{-})\met$ 
& 
\begin{tabular}{c} $\met \geqslant \, 85 \, \gev$ \\ $p_{T,\ell} \geqslant 50 \, \gev$\,,\; 
$ \big|\eta_{\ell}\big| \leqslant 2.47$ \end{tabular}
& 
\begin{tabular}{c} $\met \geqslant 178 \, \gev$ \\ $p_{T,\ell} \geqslant 100 \, \gev$ \,,\; 
$ \big|\eta_{\ell}\big| \leqslant 2.47$\end{tabular}
\\[2mm]\hline
$pp\to V(q\bar{q})\met$ 
& 
\begin{tabular}{c} $\met \geqslant \, 200 \, \gev$ \\ $p_{T,j} = 20 \, \gev$\,,\; 
$ \big|\eta_{j}\big| \leqslant 4.5$ \\ $m_{jj} \in \big[65,\, 105\big]\,\gev$ \end{tabular}
& 
\begin{tabular}{c} $\met \geqslant \, 400 \, \gev$ \\ $p_{T,j} = 40 \, \gev$ \,,\; 
$ \big|\eta_{j}\big| \leqslant 4.5$
\\ $m_{jj} \in \big[65,\, 105\big]\,\gev$ \end{tabular}
\end{tabular}
\caption{\it The upgraded configurations of LHC and the corresponding
parton-level kinematic cuts \cite{CidVidal:2018eel}.}
\label{tab:hptc}
\end{center}
\end{table}

\begin{figure}[h]
\centering
\includegraphics[width=0.44\textwidth]{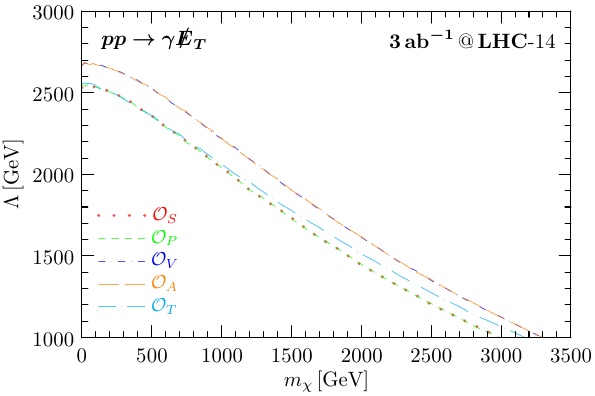}
\;
\includegraphics[width=0.44\textwidth]{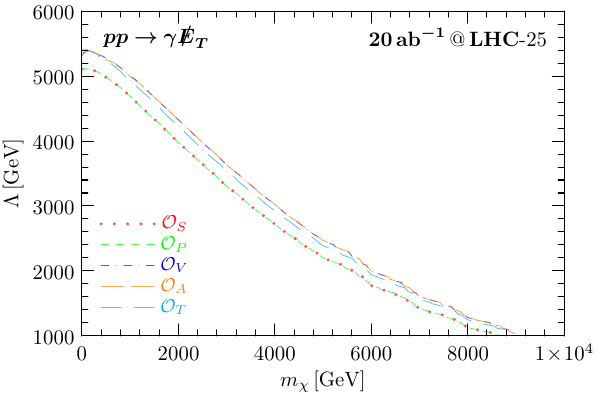}
\\
\includegraphics[width=0.44\textwidth]{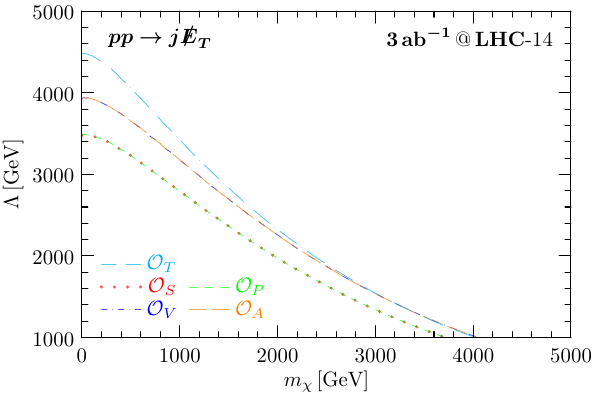}
\;
\includegraphics[width=0.44\textwidth]{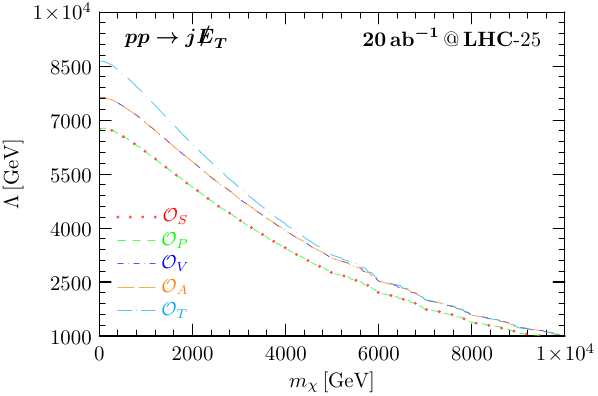}
\caption{\it 
The expected 95\%\,C.L. exclusion limits from the mono-$\gamma$
(upper) and mono-$j$ events (lower) at LHC-14 (left) and LHC-25 (right).
}
\label{fig:sens:pers:LHC}
\end{figure}
\begin{figure}[h!]
\centering
\includegraphics[width=0.44\textwidth]{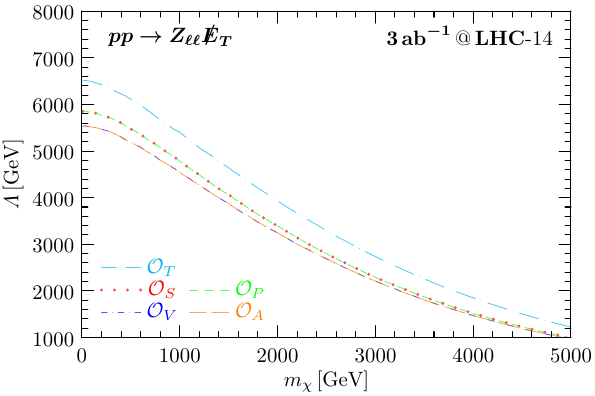}
\;
\includegraphics[width=0.44\textwidth]{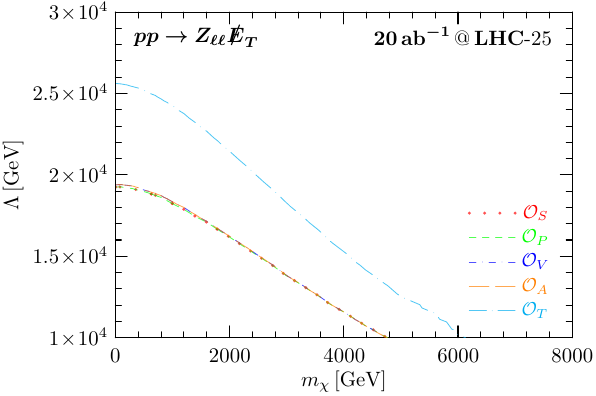}
\\
\includegraphics[width=0.44\textwidth]{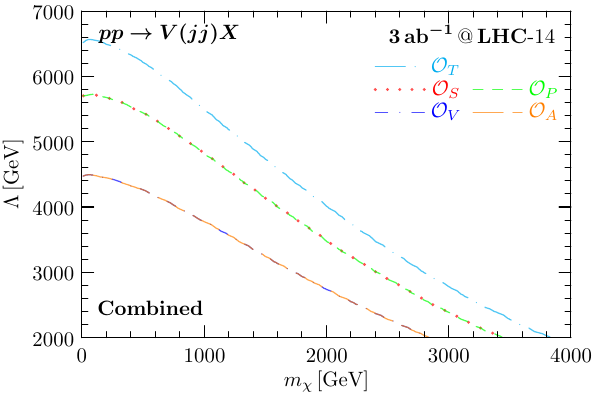}
\;
\includegraphics[width=0.44\textwidth]{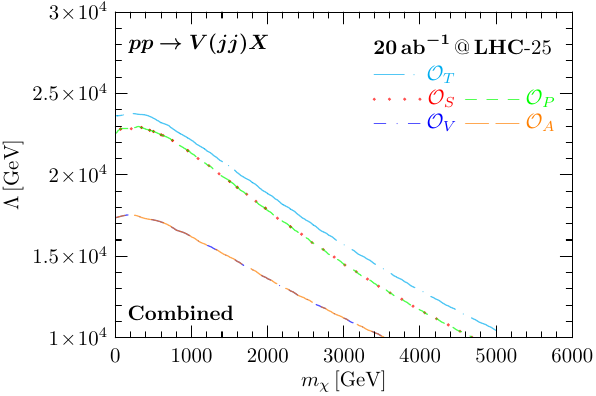}
\caption{\it 
The expected 95\%\,C.L. exclusion limits from the mono-$Z_{\ell \ell}$
(upper) and mono-$V_{jj}$ events (lower)
at LHC-14 (left) and LHC-25 (right).
}
\label{fig:sens:mz:pers:LHC}
\end{figure}
\gfig{fig:sens:pers:LHC} and \gfig{fig:sens:mz:pers:LHC}
show the expected 95\%\,C.L. exclusion limits using the mono-$\gamma$,
mono-$j$, mono-$Z_{\ell\ell}$, and mono-$V_{jj}$ (combined results of
the HP and LP regions) events at the LHC-14 (left) and LHC-25 (right),
respectively.
One can clearly see the constraint improvements, although the enhancements
vary from process to process. However, for all the processes, there is roughly 
a factor of 1.5 and 3 enhancement on the lower bounds of the energy scale 
$\Lambda_i$ (for $m_\chi=0$) at the LHC-14 and LHC-25
configurations, respectively.

\subsection{Possible Enhancements}

The sensitivities obtained above are very conservative in
two aspects. First, in the mixing picture provided by the UV
completion \cite{Dror:2019dib}, not just those absorption
operators in \geqn{eq:O} with one dark fermion $\chi$ and
one neutrino can be generated but also diagonal terms with two dark
fermions or two neutrinos. For a small mixing strength
$\Theta$, the diagonal terms dominate over the absorption
operators considered in this paper. Then one may expect much
more mono-$\gamma$, mono-jet, and mono-$Z$ events with the
missing energy taken away by a pair of dark fermions $\chi \bar \chi$
or neutrinos ($\nu \bar \nu$). Consequently, the sensitivity
can be even further improved with larger data sample.
For very light dark fermion, $m_\chi \ll \sqrt s$, with
mass $m_\chi$ being much smaller than the center-of-mass
energy $\sqrt s$, the event rates for different combinations
of final-state particles follow essentially the same spectrum
shape and can simply scale up by a factor
of $1 / \Theta^2$. Since the event rates scale with the
cut-off scale $\Lambda$ as $1 / \Lambda^4$, a naive estimation
of the sensitivity on $\Lambda$ would lead to a factor
$1/ \sqrt \Theta$ enhancement. However, we only focus
on the absorption operators in the current study to be
conservative and keep generality without too much model
assumptions.

Furthermore, it is possible for the final-state dark
fermion to have sizable decay width $\varGamma_\chi$.
Especially, it may become visible at the collider
detector if the dark fermion decays fast enough into
the SM particles. It is then possible to distinguish
the visible and invisible decay channels of the produced
dark fermion at collider. However, adding
visible decay channels would reduce the branching
ratio of the invisible channels but would not reduce
the total event rate of the mono-$\gamma$, mono-jet,
or mono-$Z$ events. Then the projected sensitivity
can only be further enhanced rather than significantly
reduced. For illustration, we consider the most
conservative case with an invisible dark fermion
in the current study.

\begin{figure}[t] 
\centering
\includegraphics[width=0.58\textwidth]{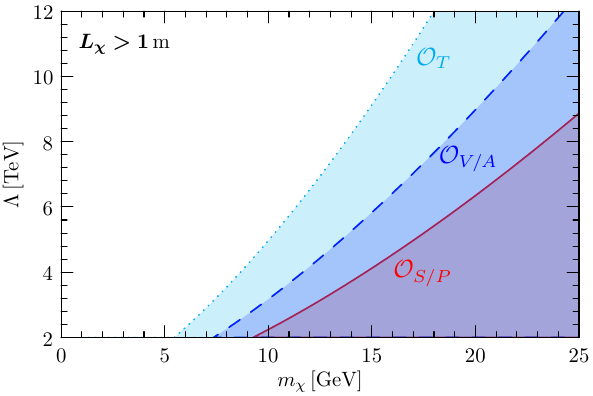}
\caption{\it 
Typical decay length of the dark fermion in the $m_\chi$-$\varLambda$ plane.
The parton level center of mass energy is chosen as $\sqrt{\hat{s}}=1\tev$.
}
\label{fig:dcylen}
\end{figure}

However, the possibility of distinguishing the visible
and invisible decay channels is an interesting
extension of our current study. Although we would
not get into too many details, it is useful to illustrate
the probable parameter space. With absorption operators
that involve quarks, the dark fermion can decay into
a neutrino plus a pair of quarks, $\chi \to \nu + q + \bar q$.
The total decay widths are given as,
\begin{align}
  \varGamma
=
\left\{
  \frac{ m_\chi^5 }{1024 \pi^3 \varLambda_{S/P}^4},
  \frac{ m_\chi^5 }{ 256 \pi^3 \varLambda_{V/A}^4},
  \frac{ 3 m_\chi^5 }{ 128 \pi^3 \varLambda_{T}^4}
\right\},
\end{align}
for the five effective operator types.
Notably, the total decay width has $m_\chi^5$ dependence. 
At a collider with a parton-level center-of-mass energy
$\sqrt{\hat{s}}$, the typical decay length of the dark
fermion is,
\bee
  L_{\chi} 
= \frac {\gamma_\chi} {\Gamma_\chi}
= \frac{ \sqrt{\hat{s}} }{2 m_\chi \varGamma_{\chi} }  
\Big( 1 + \frac{ m_\chi^2 }{ \sqrt{\hat{s}} } \Big).
\ene
For a relatively light dark fermion, this simplifies
to $L_\chi \approx \sqrt{\hat s}/2 m_\chi \varGamma_\chi$.
As a benchmark, we require the typical decay length
to exceed one meter. \gfig{fig:dcylen} shows the
regions where \(L_\chi < 1\,\text{m}\) for a typical
LHC parton-level center-of-mass energy of \(\sqrt{\hat{s}}=1\ \text{TeV}\).
As one can see, the invisibility assumption is always
valid for \(m_\chi \lesssim 5\ \text{GeV}\) which
covers the stable DM particle region with a tiny mass
$m_\chi \lesssim 100$\,MeV. Note that within this mass window, 
the direct decay channels are forbidden by the QCD
color confinement.
For a heavier dark fermion, the sensitivity should
at least remain roughly the same if not better.

\section{Absorption at Nuclear Targets}
\label{sec:ncc}

The above \gsec{sec:Bound:LHC} deals with the DS fermions
that have absorption operator in general which is a great
advantage of collider searches. The DM particle that survives
until today in our Universe is just a part of the DS particle
zoo. It is then possible to establish complementary searches
between the collider and direct detection experiments.
However, the DM particle needs to be stable enough. For
a fermionic absorption DM, its mass cannot be too large
and is typically below 100\,MeV \cite{Dror:2019onn,Dror:2019dib,Fitzpatrick:2023xks,Dror:2023fyd,Anilkumar:2024tda}.
In the current section, we would focus on this light mass
regime.
 
The parton-level operators in \geqn{eq:effo} inevitably
lead to signals at the hadron level that can be investigated by
the DM direct detection experiments. 
Here we assume that the DM particle is light enough such that its stability and relic abundance
can be easily accommodated.
The DM is converted to an active
neutrino and its mass fully absorbed to generate large nucleus
recoil energy \cite{Dror:2019onn,Dror:2019dib},
\bee
\label{eq:abs:proc:nuc}
  \chi(p_\chi) + A(p_i)
\rightarrow
  \nu(p_\nu) + A(p_f) \,,
\ene
where we have used the mass number $A$ to denote the nucleus target. 
Those momenta of the initial and final states are explicitly specified
in parenthesis.
The differential scattering rate per nuclear recoil energy $E_R$ is given as,
\bee
\label{eq:dd:drt}
\frac{\d R_A}{\d E_R}
= 
N_A n_\chi
\int \mathrm{d}^3 \bm v_\chi f_{\mathrm{E}}(\boldsymbol{v}_\chi, t) v_\chi 
\frac{\mathrm{d} \sigma_A}{\mathrm{~d} E_{\mathrm{R}}} \,, 
\ene
where $N_A$ is the number of nuclear targets. If there are more
than one isotope, 
the total rate is simply the sum of the individual contributions.
In addition,
$n_\chi \equiv  \rho_\chi/m_\chi$ is the DM local number density 
with the local DM energy density $\rho_\chi \simeq 0.3 {\,\rm GeV/cm}^3$.
The differential cross section
$\d\sigma_{\rm A}/\d E_{\rm R}$ in \geqn{eq:abs:proc:nuc}
is averaged over the incoming DM velocity distribution
$f_{\mathrm{E}}(\boldsymbol{v}_\chi, t)$.

In contrast to the usual elastic and inelastic scatterings,
the absorption process has quite different kinematics.
For a non-relativistic DM, the nucleus recoil energy,
$E^0_R = m_\chi^2 / 2 (m_A + m_\chi)$, is roughly proportional to the DM mass
squared $m^2_\chi$ at the leading order with $m_\chi$ much smaller
than the nucleus mass $m_A$. So the differential cross section 
$\d\sigma_{\rm A}/\d E_R$ exhibits a sharp peak at $E^0_R$,
\bee
\label{eq:dd:dxst}
\frac{\d\sigma_{A}}{\d E_{\rm R}}
= 
\frac{ \overline{\big| \calm_{A} \big|^2} }{16 \pi m_{A}^2 v_\chi  } 
\delta\big( E_R - E_R^0 \big)\,,
\ene
where $\overline{ | \mathcal{M}_A |^2 }$ is the squared amplitude of the process in
\geqn{eq:abs:proc:nuc} with helicity average and summation over the 
initial and final states, respectively. For a nucleus with total spin $J$,
the explicit definition of $\overline{ | {\mathcal M}_A |^2 }$ is given as,
\bee
\overline{|\mathcal{M}_A|^2} 
\equiv 
\frac{1}{2 s_{\chi}+1} \frac{1}{2 J+1} 
\sum_{\substack{\text { initial \& } \\ \text { final spins }}}|\mathcal{M}_A|^2 \,,
\ene
where $s_{\chi}=1/2$ is the DM spin. With a non-relativistic DM,
\ie, $m_\chi \gg p_\chi$, the dominant contribution to the amplitude is given by the limit
$p_\chi \to 0$. Therefore, at leading order, 
the absorption process amplitude
$\mathcal{M}_A$ is independent of the DM velocity. In this case, the velocity integral 
in \geqn{eq:dd:drt} can be carried out independently.
Inserting \geqn{eq:dd:dxst} into
\geqn{eq:dd:drt}, the DM velocity dependence can be completely removed
since the DM velocity distribution function should be normalized to 1.
In practice, there is also a threshold of the recoil energy, $E_{\rm R}^{\rm th}$, 
which is the minimal energy that can be detectable in a specific experiment,
\bee
\frac{\mathrm{d} R_A}{\mathrm{d} E_{\mathrm{R}}}
=
\frac{ N_A  n_\chi }{16 \pi m_{A}^2  } \overline{\big| \calm_A \big|^2}
\delta\big( E_R - E_R^0 \big)  \varTheta(E_{\rm R} - E_{\rm R}^{\rm th}) \,.
\label{eq:diffEventRate}
\ene

Although both the coherent and incoherent scatterings can contribute
to the direct detection of absorption DM \cite{Ge:2024euk}, we
consider only the coherent one for simplicity and easy comparison
with experimental analysis. Both the spin-independent (SI)
and spin-dependent (SD) interactions can happen depending on
the Lorentz structure of the various effective operators under study.
The nucleus-level amplitude $\calm_A$ can be written as,
\bee
\label{eq:nucleus:ff}
\calm_A(q^2) 
= 
\sum_{N=p, n} F_{\rm Res}^{N} (q^2)\calm_{\rm PLN}^{N}(q^2)
=
\sum_{N=p, n} F_{\rm Res}^{N} (q^2) C_N \calm_{N}(q^2) \,,
\ene
where $q^2 \equiv (p_f-p_i)^2 = (p_\nu - p_\chi)^2$
is the squared momentum transfer
and $F_{\rm Res}^{N}(q^2)$ are the corresponding response functions.
The amplitude for the DM scattering
off a point-like nucleus (PLN) $\calm_{\rm PLN}^{N}(q^2)$ is related
to the scattering amplitude $\calm_{N}(q^2)$ off a single nucleon
(with nucleon mass normalized to $m_A$). The difference between the
SI and SD interactions are distinguished by a coherence factor $C_N$.
For the SI scattering, $C_N = Z$ and $A-Z$ for proton and neutron,
respectively, while $C_N=1$ for its SD counterpart.

In order to caculate the DM direct detection event rate according
to \geqn{eq:diffEventRate}, the quark-level operators in \geqn{eq:effo}
should be converted to the nucleon-level matrix elements
\cite{Shifman:1978zn,Drees:1993bu,Crivellin:2013ipa,Bertuzzo:2017lwt,Bishara:2017pfq},
\begin{subequations}
\begin{align}
\label{eq:match:s}
\langle N |  m_q\bar q  q | N \rangle
&=
F_{S}^{q/N}(q^2) \overline{u}_{N} u_{N} \,,
\\[2mm]
\label{eq:match:p}
\langle N |  m_q \bar q i\gamma_5 q | N \rangle
&=
F_{P}^{q/N}(q^2) \overline{u}_{N} i\gamma_5 u_{N} \,,
\\[2mm]
\langle N |  \bar q \gamma^{\mu} q | N \rangle
&=
\label{eq:match:v}
\overline{u}_{N} \bigg[ 
F_{1}^{q/N}(q^2) \gamma^\mu  + \frac{i\sigma^{\mu\nu} q_\nu}{2m_{N}} F_{2}^{q/N}(q^2)  
\bigg] u_{N} \,,
\\[2mm]
\label{eq:match:a}
\langle N |  \bar q \gamma^{\mu}\gamma_5 q | N \rangle
&=
\overline{u}_{N} \bigg[ 
F_{A}^{q/N}(q^2) \gamma^\mu\gamma_5 + 
\frac{\gamma_5 q^\mu}{2m_{N}} F_{P'}^{q/N}(q^2)  
\bigg] u_{N} \,,
\\[2mm]
\label{eq:match:t}
\langle N |  m_q \bar q \sigma^{\mu\nu} q | N \rangle
&=
\overline{u}_{N} \bigg[ 
F_{T,0}^{q/N}(q^2) \sigma^{\mu\nu} 
+ \frac{i \gamma^{[\mu}q^{\nu]}  }{2m_{N}} F_{T,1}^{q/N}(q^2)
+
\frac{i q^{[\mu} p^{\nu]}  }{m_{N}^2} F_{T,2}^{q/N}(q^2)  
\bigg] u_{N} \,,
\end{align}
\end{subequations}
where $p \equiv p_f + p_i$ is sum of the nucleus momenta.
In general, the form factors are complex functions of the
momentum transfer squared $q^2$.
Here we consider only the leading contributions to the form factors
as given in \gapp{sec:app:nff}.

In the non-relativistic limit, the scalar and vector
nucleon bilinear operators are SI
while the pseudo-scalar, axial-vector, and tensor operators are SD.
According to \geqn{eq:match:s} and \geqn{eq:match:p},
the quark-level scalar and pseudo-scalar operators exactly 
match the scalar and pseudo-scalar operators 
at the nucleon level, respectively. Hence, the scalar operator in
\geqn{eq:oS} can induce only SI while the pseudo-scalar
operator in \geqn{eq:oP} can induce only SD scatterings.
For the quark-level vector operator, both the vector ($F^{q/N}_1 \gamma^\mu$)
and tensor ($i F^{q/N}_2 \sigma^{\mu \nu} q_\nu / 2 m_N$)
interactions can appear at the nucleon level
as shown in \geqn{eq:match:v}. Hence, the SI and SD scatterings
can simultaneously happen. However, not just the SD scattering
amplitude induced by the tensor component cannot receive the
the coherence enhancement $Z^2$ (compared with the SI contribution),
but is also suppressed by a factor of $q^\mu/m_N \to m_\chi/m_{A}$
in the absorption process. Therefore, we will neglect the SD contribution 
in our following calculation for the quark-level vector operator.
Similar thing happens for the axial-vector operator.
The situation for the tensor operator is slightly different. 
While the first term of \geqn{eq:match:t} can induce SD interaction, 
the second and third ones induces SI interactions. Although 
the SI terms are suppressed by a factor of $m_\chi/m_{A}$,
their contributions can be comparable to the SD one with the coherence 
enhancement effect for the SI scattering \cite{Liang:2024tef}. 
Both the SI and SD contributions
should be considered for the quark-level tensor operator. 

In addition, since the neutrinos in our case is always relativistic, 
the explicit expressions of the non-relativistic operators can be different.
Hence, we revisit the non-relativistic expansions of both
the nucleon pair and DM-neutrino pair bilinears
in \gapp{app:NR}.

\subsection{Spin-Independent Absorption}
\label{sec:DD:SI}

From the non-relativistic expansions given in \gapp{app:NR},
one can clearly see that the lowest order contributions of
both the scalar and vector operators are SI. 
While the spatial component of the vector operator can also induce a SD interaction,
it is suppressed by a factor $q/m_{A} \sim m_\chi/m_{A}$ compared to the
temporal counterpart.
So we investigate only the SI interactions of the vector operator. \gtab{tab:SI:Exposure} lists the experiments
studied in this paper for probing the SI absorption signals.
\begin{table}[th]
\renewcommand\arraystretch{1.58}
\begin{center}
\begin{tabular}{c|c c c c}
~~~Experiment~~~   & ~~~~~~~~Target~~~~~~~~ & ~~~~~Exposure~~~~~
& ~~~~~$E_{\rm R}^{\rm th}$~~~~~  
\\
\hline
CRESSTII \cite{CRESST:2015txj}
&
$\mathrm{Ca}\mathrm{W}\mathrm{O}_4$
& 
$52$ kg day
&
307 eV
\\
CRESST-III ($\mathrm{Ca}\mathrm{W}\mathrm{O}_4$) 
\cite{CRESST:2019jnq}
&
$\mathrm{Ca}\mathrm{W}\mathrm{O}_4$
& 
3.64 kg day
&
100 eV
\\
CRESST-III ($\mathrm{Si}$) \cite{CRESST:2022lqw}
&
$\mathrm{Si}$
& 
$55.6$ g day
&
10 eV
\\
DarkSide-50 
\cite{DarkSide:2021bnz,DarkSide-50:2022qzh,DarkSide:2022dhx,DarkSide-50:2023fcw}
&
Liquid $\mathrm{Ar}$
& 
12306 kg day
&
0.6 keV
\\
XENONnT \cite{XENON:2023cxc} 
& 
Liquid Xe
& 
1.09 t yr 
&
3 keV
\\
PandaX-4T \cite{PandaX:2023xgl} 
& 
Liquid Xe 
& 
0.55 t yr
&
3 keV
\\
Borexino \cite{BOREXINO:2023ygs}
& 
$\mathrm{C}_6 \mathrm{H}_3\left(\mathrm{CH}_3\right)_3$
& 
958.58 t yr
&
500 keV
\\
PICO-60 ($\mathrm{CF}_3 \mathrm{I}$) \cite{PICO:2023uff}
& 
$\mathrm{CF}_3 \mathrm{I}$
& 
3415 kg day 
&
20 keV
\end{tabular}
\caption{\it 
The dark matter and neutrino experiments studied here for probing the SI absorption signals.}
\label{tab:SI:Exposure}
\end{center}
\end{table}

As we have mentioned, for the SI scattering, 
the nucleus-level amplitude is simply connected to
the nucleon-level one as,
\bee
  \calm_A
=
  \sum_{N=p, n} F_{\rm SI}^{N}(\bm q^2)  C_N \calm_{N},
\ene
where the SI specific response function $F_{\rm SI}^{N}(\bm q^2)$ 
is assumed to be the same for the proton and neutron. 
It is generally given by the Helm form factor $F_{\rm Helm}(\bm q^2)$ 
\cite{Helm:1956zz, Lewin:1995rx} of the target nucleus,
\bee
\label{eq:SI:rf}
  F_{\rm SI}^{N}(\bm q^2)
=
  F_{\rm Helm}(\bm q^2)=\frac{3 j_1(|\bm q| R)}{|\bm q| R} e^{-(|\bm q| s)^2 / 2}\,,
\ene
where $j_1(x)$ is the order-1 spherical Bessel function of the first kind.
The parameters $R$ and $s$ are given as
$R \equiv \sqrt{\tilde{R}^2-5 s^2}$,
$\tilde{R}=1.2 A^{1 / 3}$\,fm,
and $s=1$\,fm \cite{Engel:1991wq,Engel:1992bf,Jungman:1995df}.
For simplicity, we assume isospin symmetry
for the nucleon-level scattering matrix elements, namely
$\calm_p = \calm_n$ and hence
$\calm_A^{\varGamma} = F_{\rm Helm}(\bm q^2) A \,\calm_{N}^{\varGamma}$
with $\varGamma = S$ and $V$ stands for Lorentz structures of the interaction operators.
The proportionality to the nucleon number $A$ is a manifestation
of the coherence enhancement.
Using the non-relativistic expansions given in \gapp{app:NR},
one can easily obtain the nucleon-level scattering matrix elements as,
\begin{subequations}
\begin{align}
\overline{|\mathcal{M}_{N}^{S}|^2}
&= 
\frac{4 m_{A}^2 m_\chi^2 }{ \varLambda^4_S } F_{S}^{N} \,,
\quad\quad
F_{S}^{N} = \sum_{q} \frac{ F^{q/N}_{S} }{m_q} \,,
\\
\overline{|\mathcal{M}_{N}^{V}|^2}
&= 
\frac{4 m_{A}^2 m_\chi^2 }{ \varLambda^4_V } F_{V}^{N}\,,
\quad\quad
F_{V}^{N} = \sum_{q} F^{q/N}_{1} \,,
\end{align}
\end{subequations}
where we have assumed that the DM couples to quarks universally.

\begin{figure}[t]
\centering
\includegraphics[width=0.49\textwidth]{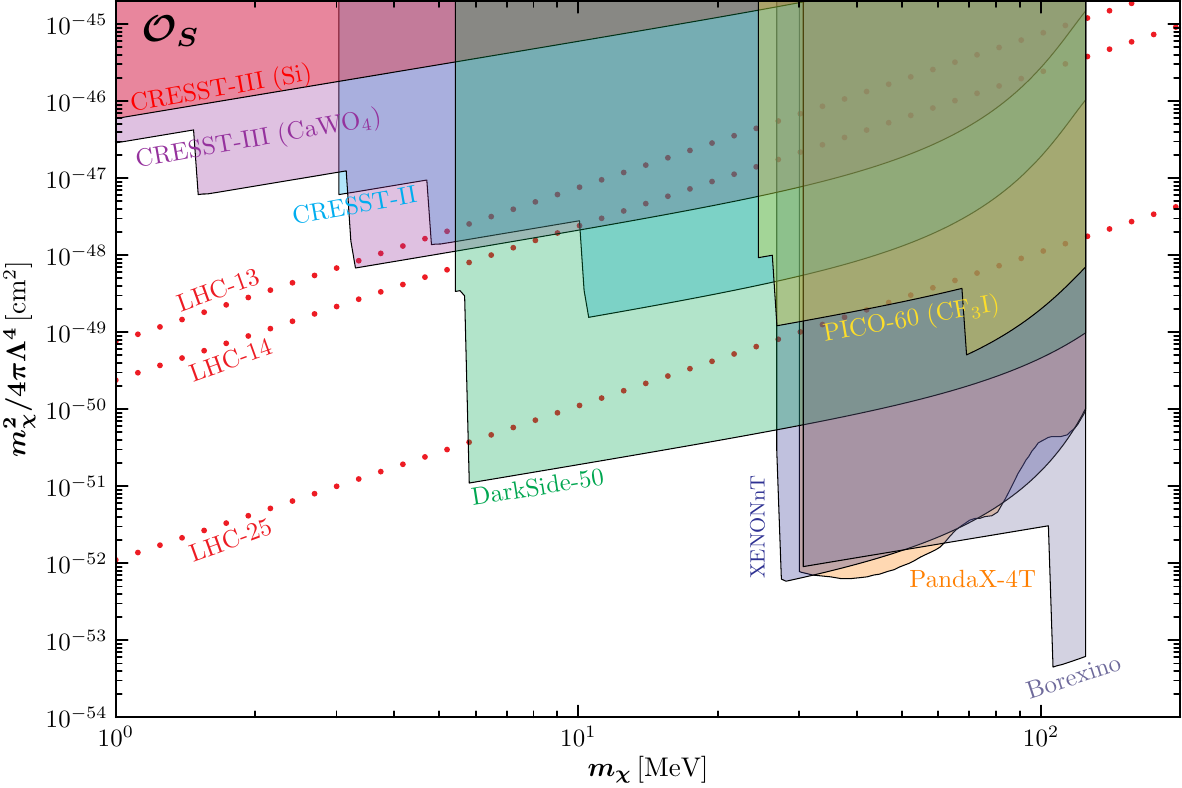}
\hfill
\includegraphics[width=0.49\textwidth]{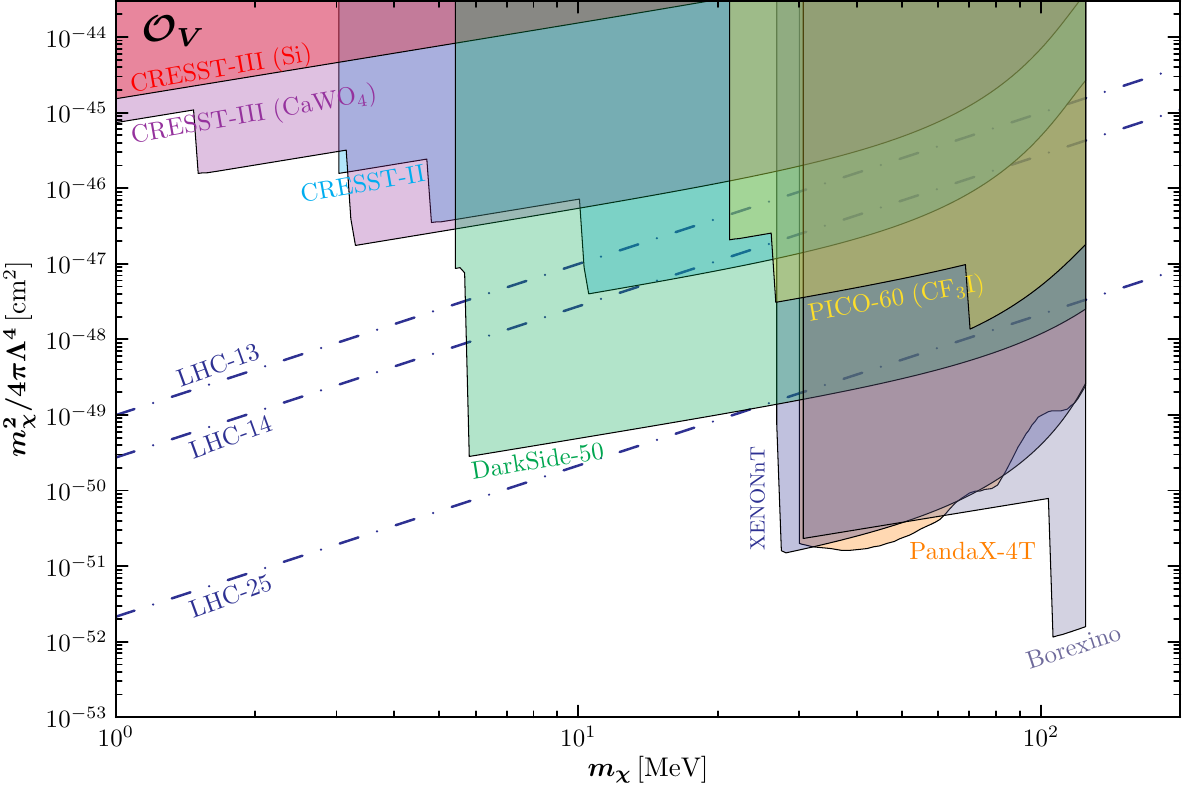}
\caption{\it 
The excluded regions on the $m_\chi^2/4\pi\varLambda^4$-$m_\chi$
plane for the scalar (\textbf{left-panel}) 
and the vector (\textbf{right-panel}) operators.
}
\label{fig:SI:CS}
\end{figure}

Putting things together, the scattering rate takes the form as follows,
\bee
R_A^{\varGamma}
=
N_A  n_\chi  \frac{ m_{\chi}^2 }{4 \pi  \varLambda_{\varGamma} } 
\big[ A F_{\varGamma}^{N}(m_\chi^2) F_{\rm Helm}(m_\chi^2) \big]^2
\varTheta(E_{\rm R}^0 - E_{\rm R}^{\rm th}) \,,
\ene
with $\varGamma = S$ or $V$. Note that we have used the approximation $\bm q^2 \approx m_\chi^2$ \cite{Ge:2024euk}.
The PandaX collaboration has searched for the absorption signal
\cite{PandaX:2022osq} which can be reinterpreted as bounds on
the scalar and vector operators. The constraints from other
experiments in \gtab{tab:SI:Exposure} are obtained by requiring
the total event number to be greater than 10.
The obtained excluded regions on the
$m_\chi^2/4\pi\varLambda^4$-$m_\chi$ plane
are shown in \gfig{fig:SI:CS}.
In the window of $m_\chi \in [\sim 10, \sim 100]\,$MeV,
the absorption process can provide stronger constraints than
the collider searches. Outside, the LHC search always
gives stronger constraints. Between the two operators,
the scalar operator receives slightly stronger bound than its
vector counterpart due to a larger nucleon form factor when
matching from the quark-level operator to the nucleon
matrix elements.

\subsection{Spin-Dependent Absorption}
\label{sec:DD:SD}
As demonstrated in \gapp{app:NR}, the pseudo-scalar and
axial-vector operators can induce SD interactions.
Different from the axial-vector operator,
the leading contribution for the pseudo-scalar operator is 
suppressed by an extra factor of $q/m_A \sim m_\chi/m_A$.
In following studies, the nucleon-level matrix elements
proportional to $|\bm q|/m_A$ will be neglected except the pseudo-scalar
operator.
With this approximation, there are only two types of SD interactions,
\begin{subequations}
\begin{align}
\label{eq:nro:ss}
\calo_{SS}
& \equiv
\big[(\xi_{h_\nu}^{\nu})^\dag \boldsymbol{s}\, \xi^{\chi}_{h_\chi} \big]
\big[(\omega_{h_{N'}}^{N'})^\dag  \boldsymbol{s} \,\omega_{h_N}^N \big] \,,
\\
\calo_{LS}
& \equiv
\big[(\xi_{h_\nu}^{\nu})^\dag \xi^{\chi}_{h_\chi} \big]
\big[(\omega_{h_{N'}}^{N'})^\dag  \big( \boldsymbol{q} \cdot \boldsymbol{s}\, \big) \,\omega_{h_N}^N \big] \,,
\end{align}
\end{subequations}
where $\xi_{h_a}^a$ are the two-component spinors with helicity $h_a$
for the neutrino ($a=v$) and the DM ($a=\chi$) while
$\omega_{h_a}^a$ are the two-component spinors with helicity $h_a$
for the incoming nucleon ($a=N$) and the outgoing nucleon ($a=N'$).
Using the non-relativistic expansions in \gapp{app:NR},
the nucleon-level operators for the pseudo-scalar
and axial-vector operators can be rewritten in terms of $\calo_{SS}$ and $\calo_{LS}$ 
as,
\begin{subequations}
\begin{align}
  \calo^{N}_{P}
& =
  2 \sqrt{2} h_{\nu} m_\chi F_{P}^{N} \calo_{LS} \,,
\quad\quad
  F_{P}^{N} = \sum_{q} \frac{ F^{q/N}_{P} }{m_q} \,,
\\
  \calo^{N}_{A}
& = 
  8 \sqrt{2}  m_\chi m_A F_{A}^{N} \calo_{SS} \,,
\quad\quad
F_{A}^{N} = \sum_{q} F^{q/N}_{A} \,,
\end{align}
\end{subequations}
where $F_{P}^{N}$ and $F_{A}^{N}$ are the nucleon form factors
for the pseudo-scalar and axial-vector operators, respectively.
At the nucleus level, the matrix elements of the non-relativistic operators
 are,
\begin{subequations}
\begin{align}
  \calm_{A,SS}
& =
  \sum_{N=p, n}
  \boldsymbol{s}_\chi \cdot \left\langle J, M'\left|\boldsymbol{S}_N\right| J, M\right\rangle,
\label{eq:MASS}
\\
  \calm_{A, LS}
& =
  \sum_{N=p, n} \mathbb{I}_\chi \;
  \big[ \left\langle J, M^{\prime}\left|\boldsymbol{S}_N\right| J, M\right\rangle \cdot \boldsymbol{q} \big],
\end{align}
\end{subequations}
where $\boldsymbol{s}_\chi$ is the DM/neutrino spin operators, 
and $ \mathbb{I}_\chi $ is an identity operator in the spinor space.
While $J$ and $M$ are the total and projected spin quantum numbers,
$\boldsymbol{S}_N$ is the total nucleon spin operator
defined as $\boldsymbol{S}_p \equiv \sum_{i=\text { protons }} \boldsymbol{s}_{p_i}$
and $\boldsymbol{S}_n \equiv \sum_{i=\text { neutrons }} \boldsymbol{s}_{n_i}$
for proton and neutron, respectively.
For vanishing total spin $J=0$, the above matrix elements clearly vanishes. 
Some typical experiments with isotopes having non-zero total spin,
are listed in \gtab{tab:SD:Exposure}.
We will study the SD signals at these experiments.
\begin{table}[t]
\renewcommand\arraystretch{1.58}
\begin{center}
\begin{tabular}{c|c c c c c}
Experiment  & Target & Exposure & Isotope (Abund.) & $E_{\rm R}^{\rm th}$
\\
\hline\hline
XENONnT \cite{XENON:2023cxc}
& 
Liquid Xe 
& 
1.09\,t$\times$yr 
&
\begin{tabular}{c} 
$ {}^{129}_{\;\;54}\mathrm{Xe} $ (26.4\%) \\ 
$ {}^{131}_{\;\;54}\mathrm{Xe} $ (21.2\%) 
\end{tabular}
& 
3\,keV
\\
\hline
PandaX-4T \cite{PandaX:2022xas}
&
Liquid Xe
& 
0.63\,t$\times$yr 
& 
\begin{tabular}{c} 
$ {}^{129}_{\;\;54}\mathrm{Xe} $ (26.4\%) \\ 
$ {}^{131}_{\;\;54}\mathrm{Xe} $ (21.2\%) 
\end{tabular}
& 
3\,keV
\\
\hline
Borexino \cite{BOREXINO:2023ygs} 
& 
$\mathrm{C}_6 \mathrm{H}_3\left(\mathrm{CH}_3\right)_3$
& 
958.58\,t$\times$yr
&
\begin{tabular}{c} 
$ {}^{13}_{\;\,6}\mathrm{C} $ (1.1\%) \\
$ {}^{1}_{1}\mathrm{H} $ (99.985\%) 
\end{tabular}
&
500\,keV
\\
\hline
CRESST-III ($\mathrm{Li}\mathrm{Al}\mathrm{O}_2$) \cite{CRESST:2022dtl}
&
$\mathrm{Li}\mathrm{Al}\mathrm{O}_2$
& 
2.345\,kg$\times$day 
&
\begin{tabular}{c} 
$ {}^{6}_{3}\mathrm{Li} $ (7.5\%) \\ $ {}^{7}_{3}\mathrm{Li} $ (92.5\%)
\\ $ {}^{27}_{13}\mathrm{Al} $ (100\%) 
\end{tabular}
& 
94.1\,eV
\\
\hline
PICO-60 ($\mathrm{C}_3 \mathrm{F}_8$) \cite{PICO:2023uff} 
& 
$\mathrm{C}_3 \mathrm{F}_8$
& 
2571\,kg$\times$day 
&
\begin{tabular}{c} 
$ {}^{13}_{\;\,6}\mathrm{C} $ (1.1\%) \\
$ {}^{19}_{\;\,9}\mathrm{F} $ (100\%) 
\end{tabular}
&
3.3\,keV
\end{tabular}
\caption{\it 
The dark matter and neutrino experiments studied here for probing the SD absorption signals.
}
\label{tab:SD:Exposure}
\end{center}
\end{table}

The squared amplitudes summing over the final-state spins
and averaging over the initial states are given as,
\begin{subequations}
\begin{align}
\label{eq:ss:sm}
  \overline{\big| \calm_{A, SS} \big|^2}
& =
  \frac{1}{2} \sum_{N, N'} F_{4,4}^{NN'} 
=
  \sum_{N, N'}  \frac{1}{32} \Big[ F_{\Sigma'}^{NN'}(q^2) + F_{\Sigma''}^{NN'}(q^2)  \Big],
\\
\label{eq:ls:sm}
  \overline{\big| \calm_{A,LS} \big|^2}
& =
  \sum_{N, N'} F_{10,10}^{NN'} 
=
  \sum_{N, N'}  \frac{q^2}{4} F_{\Sigma''}^{NN'}(q^2).
\end{align}
\label{eq:MA}
\end{subequations}
There is an additional factor $1/2$ due to the fact that 
only the left-handed neutrino can participate the scattering.
The response functions are defined as,
\begin{subequations}
\begin{align}
\label{eq:ss:sm}
F_{4,4}^{NN'} 
& \equiv
\frac{1}{4(2J+1)} \sum_{M, M'} 
\left\langle J, M\left|\boldsymbol{S}_{N'}\right| J, M'\right\rangle
\left\langle J, M'\left|\boldsymbol{S}_N\right| J, M\right\rangle \,,
\\
\label{eq:ls:sm}
 F_{10,10}^{NN'} 
& \equiv
\frac{1}{2J+1}
\sum_{M, M'} 
\left\langle J, M\left|\boldsymbol{S}_{N'} \cdot \boldsymbol{q} \right| J, M'\right\rangle
\left\langle J, M'\left|\boldsymbol{S}_N \cdot \boldsymbol{q} \right| J, M\right\rangle \,.
\end{align}
\end{subequations}
For the non-relativistic operator $\calo_{SS}$ in
\geqn{eq:nro:ss}, the nuclear response function
can be of both the $\Sigma^{'}$ and $\Sigma^{''}$ types as
defined in \geqn{eq:MA},
while for the operator $\calo_{LS}$, there is only the $\Sigma^{''}$ response  \cite{DelNobile:2021wmp}.
In general, these response functions are complex functions of the momentum
transfer $q$ and model parameters. 
The explicit form of 
$F_{\Sigma'}^{NN'}$ and $F_{\Sigma''}^{NN'}$ for some isotopes can be found
in \cite{Fitzpatrick:2012ix}.
Here we simply employ the zero momentum transfer approximation
where the response function $F_{10,10}^{NN'} \propto \bm q^2$ vanishes. 
However, with the following relation at the zero momentum transfer,
\bee
\label{eq:ssr}
 \sum_{M, M'} 
\left\langle J, M\left|S_{N', i}\right| J, M'\right\rangle
\left\langle J, M'\left|S_{N, j}\right| J, M\right\rangle
=
\frac{(J+1)(2 J+1)}{3 J} 
\mathbb{S}_N \mathbb{S}_{N'} \delta_{i j} \,,
\ene
where $\mathbb{S}_N \equiv \left\langle J, J \left|S_N^z\right| J, J\right\rangle$ 
are the expectation values of the nucleus spin operator 
for the maximal spin projection,
the response functions are given as,
\begin{subequations}
\begin{align}
\label{eq:ss:boundary}
F_{4,4}^{NN'} 
&=
\frac{J+1}{4J} \mathbb{S}_N \mathbb{S}_{N'}  \,,
\quad\quad
F_{\Sigma'}^{NN'} + F_{\Sigma''}^{NN'} = \frac{4(J+1)}{J}  \mathbb{S}_N \mathbb{S}_{N'} \,,
\\
F_{10,10}^{NN'} 
&=
\frac{J+1}{3J} \bm q^2 \mathbb{S}_N \mathbb{S}_{N'}  \,,
\quad\quad
F_{\Sigma''}^{NN'} = \frac{4(J+1)}{3J}  \mathbb{S}_N \mathbb{S}_{N'} \,.
\end{align}
\end{subequations}
Particularly, the response function $F_{10,10}^{NN'}(\bm q^2)$ is
approximately $F_{10,10}^{NN'}(\bm q^2=0) \approx \bm q^2/4\, F_{\Sigma''}^{NN'}(\bm q^2=0)$
such that the $\bm q^2$ dependence is factorized out.
Since determining $\mathbb{S}_N$ requires detailed calculations 
within realistic nuclear models, its values 
found in the literature are often model dependent and 
sometimes differ from each other for a given nucleus. 
\gtab{tab:exps} summarizes the spin expectation values 
($\mathbb{S}_N$) of some isotopes that will be studied in this paper.
\begin{table}[t]
\renewcommand\arraystretch{1.58}
\begin{center}
\begin{tabular}{c|c c c c c}
Isotope (Abund.) & $J$ & $\mathbb{S}_{p}$ & $\mathbb{S}_{n}$ & Ref.
\\\hline\hline
$ {}^{1}_{1}\mathrm{H} $ (99.985\%)  & 1/2 & 0.5 & 0 & \cite{Ellis:1987sh,Ellis:1991ef}
\\[2mm]\hline
$ {}^{6}_{3}\mathrm{Li} $ (7.5\%)  & 1/2 & 0.472 & 0.472 & \cite{Girlanda:2011fh,CRESST:2022dtl}
\\
$ {}^{7}_{3}\mathrm{Li} $ (92.5\%) & 3/2 & 0.497 & 0.004 & \cite{Pacheco:1989jz}
\\[2mm]\hline
$ {}^{13}_{\,\,\,6}\mathrm{C} $ (1.1\%)  & 1/2 & $-0.009$ & $-0.172$ & \cite{Engel:1989ix}
\\[2mm]\hline
$ {}^{19}_{\,\,\,9}\mathrm{F} $ (100\%)  & 1/2 & 0.475 & $-0.0087$ & \cite{Divari:2000dc}
\\[2mm]\hline
$ {}^{27}_{13}\mathrm{Al} $ (100\%)   & 5/2 & 0.343 & 0.0296 & \cite{Engel:1995gw}
\\[2mm]\hline
\begin{tabular}{c} 
$ {}^{129}_{\;\;54}\mathrm{Xe} $ (26.4\%) \\ 
$ {}^{131}_{\;\;54}\mathrm{Xe} $ (21.2\%) 
\end{tabular}
& 
\begin{tabular}{r} 1/2 \\  3/2  \end{tabular}
& 
\begin{tabular}{r} 0.0128 \\  $-0.012$  \end{tabular}
& 
\begin{tabular}{r} 0.300 \\   $-0.217$ \end{tabular}
&
\cite{Ressell:1997kx}
\end{tabular}
\caption{\it 
The spin expectation values ($\mathbb{S}_N$) of the isotopes studied in this paper.
}
\label{tab:exps}
\end{center}
\end{table}
The readers can also find some averaged values from
\cite{DelNobile:2021wmp}.
In the scattering rate calculation, it usually assumes that either proton or neutron enters the interaction. Here we consider only the nucleon that has the largest 
spin expectation value for the given isotopes in \gtab{tab:exps}.

Including the phase space factor,
the scattering rate can be written in terms of the response functions as, 
\begin{subequations}
\begin{align}
\label{eq:SD:rate:p}
R_A^{P}
&=
\frac{ N_A  n_\chi  m_{\chi}^4 }{8 \pi  \varLambda_{P}^4 m_{A}^2 } 
\varTheta(E_{\rm R}^0 - E_{\rm R}^{\rm th})
\sum_{N,N'=p, n} F_{P}^{N} F_{P}^{N'} \big[ F_{\Sigma''}^{NN'} \big]  \,,
\\
R_A^{A}
&=
\frac{ N_A  n_\chi  m_{\chi}^2 }{4 \pi  \varLambda_{A}^4  } 
\varTheta(E_{\rm R}^0 - E_{\rm R}^{\rm th})
\sum_{N,N'=p, n} F_{A}^{N} F_{A}^{N'} 
\big[ F_{\Sigma'}^{NN'} + F_{\Sigma''}^{NN'} \big] \,.
\end{align}
\end{subequations}
The SD scattering are parameterized by using the spin
structure function $S(\bm q^2) \equiv 4(2J+1)F_{4,4}^{NN'}$. 
However, the usual spin structure function $S(\bm q^2)$ is somehow
specific for the axial-vector matrix operator. For instance, 
the differential scattering rate for the axial-vector operator can 
cast into a usual form in term of $S(\bm q^2)$ (apart from the phase space factors),
\bee
\frac{\mathrm{d} R_T^{A}}{\mathrm{d} E_{\mathrm{R}}}
\propto
\frac{ N_T  n_\chi  m_{\chi}^2 }{\pi  (2J+1)\varLambda_{A}^4  } 
F_{A}^{N} F_{A}^{N'} S(\bm q^2)
=
\frac{ N_T  n_\chi  m_{\chi}^2 }{\pi \varLambda_{A}^4  } \frac{J+1}{J}
F_{A}^{N} F_{A}^{N'}\mathbb{S}_N \mathbb{S}_{N'} \frac{S(\bm q^2)}{S(0) } \,,
\ene
where the factor $S(0) =\mathbb{S}_N \mathbb{S}_{N'}  (2J+1)(J+1)/J$
is obtained 
using the $S(\bm q^2)$ given in \geqn{eq:ss:boundary}.
Similar expressions can also be found for the tensor operator.
In our case, it is better to use the response functions $F_{\Sigma'}^{NN'}$
and $F_{\Sigma''}^{NN'}$ (in the limit $\bm q^2 \to 0$), 
such that the pseudo-scalar operator contribution
can also be studied in terms of the same response functions.

\begin{figure}[t]
\centering
\includegraphics[width=0.49\textwidth]{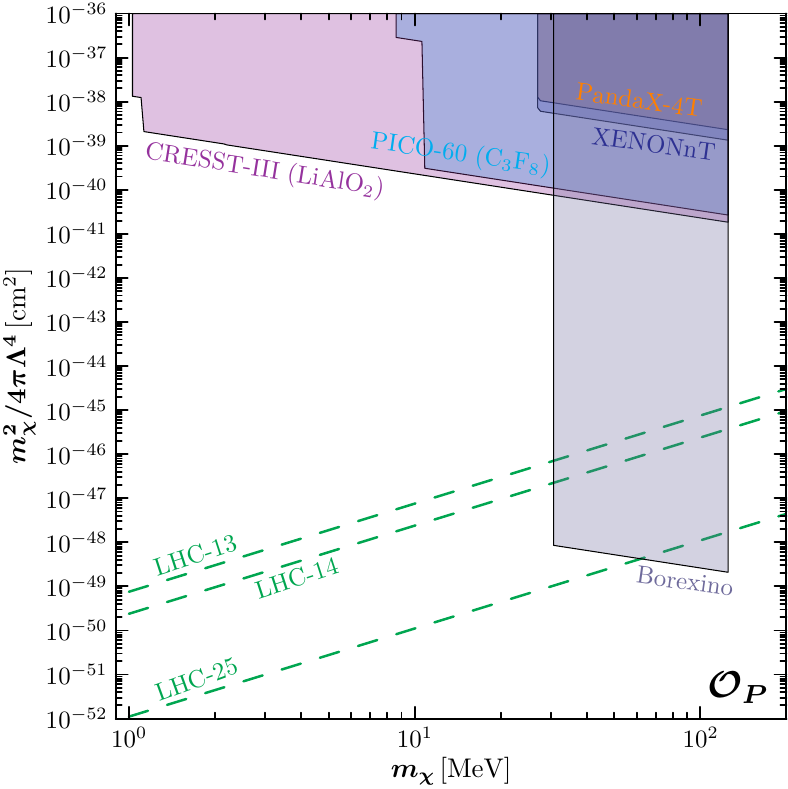}
\hfill
\includegraphics[width=0.49\textwidth]{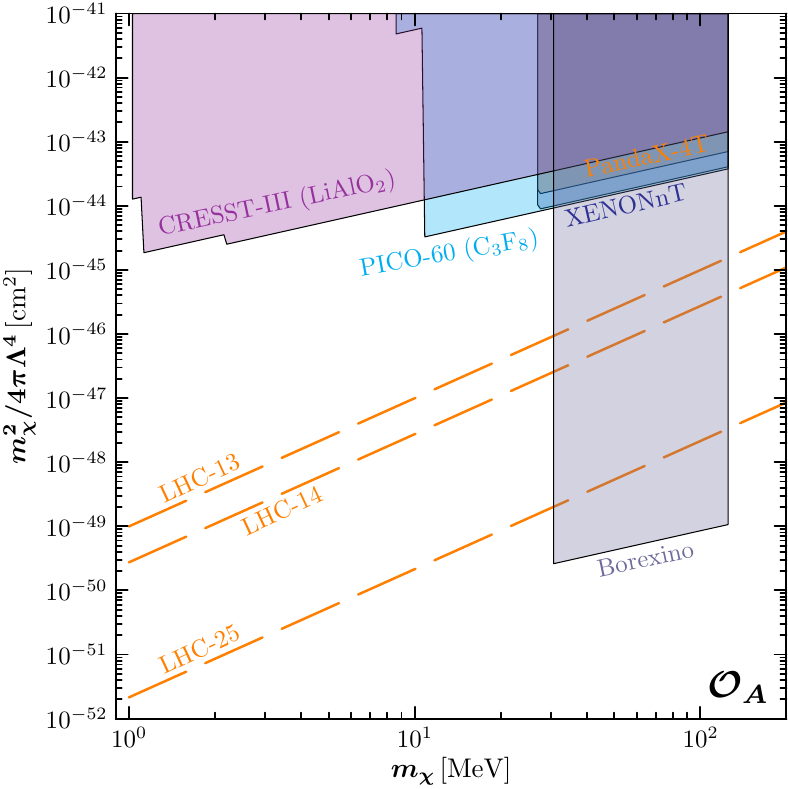}
\caption{\it 
The excluded regions on the $m_\chi^2/4\pi\varLambda^4$-$m_\chi$ plane
with SD scatterings for the pseudo-scalar (\textbf{left-panel}) 
and the axial-vector (\textbf{right-panel}) operators.
}
\label{fig:SD:CS}
\end{figure}
The exclusion bounds for those experiments listed in
\gtab{tab:SD:Exposure} are estimated by requiring the
total event number to be greater than 10.
\gfig{fig:SD:CS} shows the expected excluded regions 
on the $m_\chi^2/4\pi\varLambda^4$-$m_\chi$ plane.
Those constraints given by the SD experiments with
relatively heavier nucleus are much weaker than the
LHC searches while the light nucleus target can provide
stronger constraints. This is particularly true for
the Borexino experiment with hydrogen with two reasons.
First, the Borexino experiment has about 3 orders larger
exposure as shown in \gtab{tab:SD:Exposure}.
Second, for given detector mass,
the nuclei (proton for the Borexino experiment) number
is more than 2 orders larger than those experiments with heavy nuclei (for instance
$\mathrm{Xe}$ of PandaX and XENON).
Without coherence enhancement, the SD cross section is
roughly the same among the light and heavy nucleus and
hence the nuclei number determines the total event rate.
With these two advantages, one can see a roughly 6 orders
stronger constraint on the axial-vector operator from
Borexino (the right-panel of \gfig{fig:SD:CS}).
For the pseudo-scalar operator (the left-panel of \gfig{fig:SD:CS}), 
the enhancement at the Borexino experiment is even stronger 
since the scattering rate is suppressed by a factor $m_\chi^2/m_A^2$
which is inversely proportional to $m^2_A$. The effect
of this suppression factor also manifests itself as
stronger constraint for heavier DM in the right-panel of
\gfig{fig:SD:CS}, in contrast to the axial-vector case.
Another interesting point is that the SD constraints
can be as strong as the SI one with light nucleus
since the coherence enhancement for the SI scattering
is then dramatically diminished.

\subsection{Absorption for the Tensor Operator}
\label{sec:DD:tensor}

As mentioned easier, the quark-level tensor operator can induce 
both SI and SD interactions at the nucleon level.
So the total amplitude is the sum of two contributions,
\bee
\calm_{A,T} 
= 
\calm_{A,T}^{SD} + \calm_{A,T}^{SI} \,.
\ene

For the SD interaction induced by the first term of
\geqn{eq:match:t}, the nucleon matrix elements are dominated
by the non-relativistic operator $\calo_{SS}$ in \geqn{eq:nro:ss}
according to the non-relativistic expansions in \gapp{app:NR}, 
Similar to the axial-vector case in \geqn{eq:MASS}, the
scattering amplitude is given as,
\bee
\calm_{A,T}^{SD} =
16\sqrt{2}  m_\chi m_A \sum_{N=p,n} F_{T,SD}^{N} 
\big[ \boldsymbol{s}_\chi \cdot \left\langle J, M'\left|\boldsymbol{S}_N\right| J, M\right\rangle \big]\,,
\quad
F_{T,SD}^{N} = \sum_{q} \frac{ F^{q/N}_{T,0} }{m_q} \,.
\ene

For the SI interaction induced by the second and third terms
of \geqn{eq:match:t}, the largest contribution is given by
the off-diagonal parts of the nucleon-level tensor matrix
elements. The corresponding non-relativistic operator is,
\bee
\calo_{L} 
\equiv
\big[h_\nu(\xi_{h_\nu}^{\nu})^\dag(\boldsymbol{s}_\chi \cdot \boldsymbol{q})\, \xi^{\chi}_{h_\chi} \big] \mathbb{I}_N\,.
\ene
and the scattering amplitude is,
\bee
\calm_{A,T}^{SI} =
4 \sqrt{2} m_\chi  \sum_{N=p,n} F_{T,SI}^{N} C_N F_{SI}^{N}
\big[ \boldsymbol{s}_\chi \cdot \boldsymbol{q} \big] \delta_{M'M}\,,
\quad
F_{T,SI}^{N} = \sum_{q} \frac{ F^{q/N}_{T,1} + 2F^{q/N}_{T,2} }{m_q} \,,
\ene
where $F_{SI}^{N}$ is the SI response function defined in \geqn{eq:SI:rf}.
With two contributions to the scattering amplitude, the
total average squared amplitude,
\bee
\overline{\big| \calm_{T, L} \big|^2}
=
\overline{\big| \calm_{T, SD} \big|^2}
+
\overline{\big| \calm_{T, SI} \big|^2}
+
\overline{\calm_{T, SD}^\dag  \calm_{T, SI} + \calm_{T, SI}^\dag  \calm_{T, SD}  } \,,
\ene
contains an interference term between the SI and SD interactions.
Summing over the DM and neutrino helicity states, 
the interference can be reformed as,
\bee
\overline{\Re\big\{ \calm_{T, SD}^\dag  \calm_{T, SI} \big\} }
\propto
\sum_{M} 
\left\langle J, M \left| \boldsymbol{q} \cdot \boldsymbol{S}_N\right| J, M\right\rangle \,.
\ene
Since the nuclear target is unpolarized and the momentum
transfer orientation (or equivalently the DM velocity) is
isotropic, this interference term would not leave any effect
in the total event rate. 
\begin{figure}[t]
\centering
\includegraphics[width=0.49\textwidth]{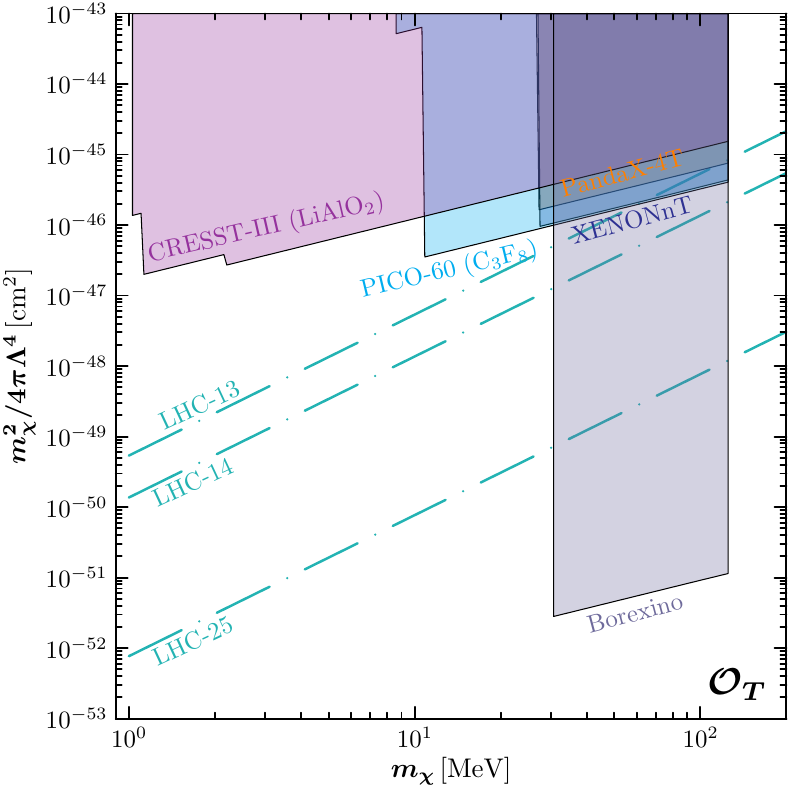}
\caption{\it 
The excluded regions on the $m_\chi^2/4\pi\varLambda^4$-$m_\chi$ plane
with SD scatterings for the tensor operator $\mathcal O_T$.
}
\label{fig:SD:T}
\end{figure}
So the total scattering rate is
just incohenrent sum of the SD and SI contributions,
\bee
R_A^{T} = R_{A, SD}^{T} + R_{A, SI}^{T} \,,
\ene
where the SD and SI scattering rates are given as follows, 
\begin{subequations}
\begin{align}
R_{A, SD}^{T} 
&=
\frac{ 16 N_A  n_\chi  m_{\chi}^2 }{\pi  \varLambda_{T}^4  } 
\varTheta(E_{\rm R}^0 - E_{\rm R}^{\rm th})
\sum_{N,N'=p, n} F_{T,SD}^{N} F_{T,SD}^{N'} 
\big[ F_{\Sigma'}^{NN'} + F_{\Sigma''}^{NN'} \big]  \,,
\\ 
R_{A, SI}^{T}
&=
N_A  n_\chi  \frac{ m_{\chi}^4 }{4 \pi  \varLambda_{T} m_{A}^2 } 
\big[ A F_{T,SI}^{N} F_{\rm Helm}(m_\chi^2) \big]^2
\varTheta(E_{\rm R}^0 - E_{\rm R}^{\rm th}) \,.
\end{align}
\end{subequations}
The SI contribution is suppressed by a factor of $m_\chi^2/m_A^2$,
similar to the pseudo-scalar contribution to the SD scattering
 in \geqn{eq:SD:rate:p}).
However, the SI contribution of the tensor operator still has
the coherency enhancement factor $A^2$ which is large for
heavier nucleus. Comparing with the SD contribution,
the SI scattering rate is roughly scaled by a factor of
$A^2 m_\chi^2/m_A^2 \approx m_\chi^2/m_N^2$ with the nucleus
mass $m_A$ replaced by the nucleon mass $m_N$. In other words,
the SI scattering on a light nuclear target needs not to be
relatively much smaller. 
However, for the DM mass window
$m_\chi \lesssim 100\,$MeV at the direct detection experiments,
the net enhancement is negligible with $m_\chi \ll m_N$.
\gfig{fig:SD:T} shows the expected excluded regions for the
tensor operator
on the $m_\chi^2/4\pi\varLambda^4$-$m_\chi$ plane.
The current LHC bound has covered almost all the constraints
given by the SD experiments with heavy nucleus. 
However, the bound obtained with light nucleus at the Borexino
experiment is still the most promising channel of detecting the
absorption signal.

\section{Conclusion}
\label{sec:conclusion}

Starting from the quark-level fermionic absorption operators,
we study for the first time their sensitivities at the LHC
and compare with the constraints from the DM direct detection
as well as neutrino experiments. While the LHC can probe the
DS particles in general, the collider and direct detection
experiments can provide complementary searches for the DM
particle that survives until today with small enough mass.
The sensitivity at LHC can reach
$\Lambda \gtrsim \mathcal O(1)$\,TeV with mono-$X$ (photon,
jet, and $Z$) channels.
Even for DM with typical mass
below $\mathcal O(100)$\,MeV, the LHC searches can provide
better sensitivities although the identification still requires
direct detection experiments that has unique recoil energy
spectrum. While for scalar and vector operators,
the DM direct detection
experiments with heavy nuclei target has much better sensitivities,
the remaining pseudo-scalar, axial-vector, and tensor operators
has much better sensitivities at neutrino experiments with light
nuclei. It is also interesting to observe that the tensor operator
can have comparable contributions from both the SI and SD
scattering.

\section*{Acknowledgements}

SFG and KM would like to thank Xiao-Dong Ma for useful discussions.
The authors are supported by the National Natural Science
Foundation of China (12375101, 12425506, 12090060, 12090064, and 12305110) and
the SJTU Double First Class start-up fund (WF220442604).
SFG is also an affiliate member of Kavli IPMU, University of Tokyo.

\appendix

\section{Nucleon Form Factors}
\label{sec:app:nff}

For completeness we list all the nucleon form factors
needed in our calculations of the DM-nucleon scattering cross sections.
We follow the parameterizations in Ref.\,\cite{Bishara:2017pfq}
and study only the leading-order corrections at the zero momentum
transfer limit $q^2 \approx 0$.
Imposing the isospin symmetry, most of the form factors 
$F_{\varGamma,\ldots}^{q/N}$ are given for proton, \ie, $N=p$,
while the neutron ones can be obtained by exchanging 
$p\to n$ and $u \leftrightarrow d$. The small differences between 
$F_{\varGamma,\ldots}^{q/p}$ and $F_{\varGamma,\ldots}^{q/n}$ 
can then be ignored.

\subsection{Scalar Current}

The scalar form factors $F_{S}^{q/N}$ are conventionally referred
as the nuclear sigma terms.
In the zero momentum transfer approximation,
\bee
F_{S}^{q/N}(q^2=0) = \sigma_{q}^{N} \,.
\ene
The matrix elements of the $u$ and $d$ quarks are related to the pion-nucleon sigma term,
$\sigma_{\pi N} \equiv \langle N | \overline{m}(\bar u u + \bar d d) |N\rangle$ with $\overline{m} \equiv (m_u+m_d)/2$.
There are several methods to obtain the experimental fits of $\sigma_{\pi N}$,
for example $\sigma_{\pi N}= (50 \pm 15)$\,MeV \cite{Bishara:2017pfq}.
And the corresponding sigma terms are \cite{Crivellin:2013ipa,Bishara:2017pfq},
\begin{equation}
\begin{aligned} 
\sigma_{u}^{p} = (17 \pm 5)\, \mathrm{MeV}\,,\quad
& 
\sigma_{d}^{p}=(32 \pm 10)\, \mathrm{MeV} \,,
\\ 
\sigma_{u}^{n}=(15 \pm 5)\, \mathrm{MeV}\,,\quad 
& 
\sigma_{d}^{n}=(36 \pm 10) \, \mathrm{MeV} \,.
\end{aligned}
\end{equation}
For strange quark, the average of lattice QCD determinations gives
\cite{Junnarkar:2013ac,Yang:2015uis,Durr:2015dna,Bishara:2017pfq},
\begin{equation}
\sigma_{s}^{p}=\sigma_{s}^{n}=(41.3 \pm 7.7) \,\mathrm{MeV} \,.
\end{equation}
The matching between the quark- and nucleon-level matrix elements
requires also inputting the quark masses
($m_u = (2.14 \pm 8)$\,MeV, $m_d = (4.70 \pm 5)$\,MeV, and
$m_{s} / \bar{m} = 27.5 \pm 0.3$)
since the matching coefficients are proportional to 
$F_{S}^{q/N}(0)/m_q$. Then we can find
\bee
\left[ \sum_{q=u,d,s} \frac{F_{S}^{q/N}(q^2=0)}{m_q} \right]^{2} =
\begin{cases}
15.2 & N=p, 
\\[1mm]
15.1 & N=n.
\end{cases} 
\ene
The difference between the matching coefficients for proton and
neutron is negligible.

\subsection{Pseudo-Scalar Current}

The pseudo-scalar current can receive contributions from
the light pseudo-scalar meson exchange and hence its amplitude
has pole structures at the meson mass points \cite{Bishara:2017pfq}.
For small momenta exchange, the form factors expands as,
\begin{equation}
F_{P}^{q / N}\left(q^{2}\right)
=
\frac{m_{N}^{2}}{m_{\pi}^{2}-q^{2}} a_{P, \pi}^{q / N}
+
\frac{m_{N}^{2}}{m_{\eta}^{2}-q^{2}} a_{P, \eta}^{q / N}
+ \cdots \,.
\end{equation}
The residua of the poles are given by
\begin{subequations}
\begin{align}
\frac{a_{P, \pi}^{u / p}}{m_{u}}
& =
-\frac{a_{P, \pi}^{d / p}}{m_{d}}
=
\frac{B_{0}}{m_{N}} g_{A}
\,, \quad 
\frac{a_{P, \pi}^{s / p}}{m_{s}}=0 \,,
\\ 
  \frac{a_{P, \eta}^{u / p}}{m_{u}}
& =
\frac{a_{P, \eta}^{d / p}}{m_{d}}
=
-\frac{1}{2} \frac{a_{P, \eta}^{s / p}}{m_{s}}
=
\frac{B_{0}}{3 m_{N}}\left(\Delta u_{p}+\Delta d_{p}-2 \Delta s_{p}\right)\,.
\end{align}
\end{subequations}
The coefficients for the neutron case can be obtained through the replacement
$p\to n$ and $u \leftrightarrow d$. In the isospin limit,
\begin{equation}
\Delta u_{n} = \Delta d_{p} \equiv \Delta u 
\,, \quad 
\Delta u_{n} = \Delta d_{p} \equiv \Delta d 
\,, \quad 
\Delta s_{n} = \Delta s_{p}  \equiv  \Delta s  \,.
\end{equation}
The iso-vector combination can be precisely determined 
from the nuclear $\beta$ decay \cite{ParticleDataGroup:2016lqr},
\begin{equation}
  \Delta u - \Delta d = g_{A}=1.2723(23).
\end{equation}
In the \(\overline{\mathrm{MS}}\) scheme at \(Q=2\,\mathrm{GeV}\), the averages of 
lattice QCD results give $\Delta u+\Delta d= 0.521(53)$ \cite{GrillidiCortona:2015jxo} and
$\Delta s=-0.031(5)$. The combination of these results gives \cite{GrillidiCortona:2015jxo},
\begin{equation}
  \Delta u=0.897(27),
\quad
  \Delta d=-0.376(27),
\quad
  \Delta s=-0.031(5).
\end{equation}
Moreover, $B_{0}$ is a ChPT (chiral perturbation theory)
constant related to the quark condensate 
and up to the order \(\mathcal{O}\left(m_{q}\right)\) 
it is given by \(\langle\bar{q} q\rangle \simeq-f^{2} B_{0}\). 
At the leading order, the quark condensate gives \(f=f_{\pi}\), with
\(f_{\pi}\) being the pion decay constant \cite{McNeile:2012xh}.
Then one has $B_{0}=2.666(57)\,\mathrm{GeV}$ at the scale $\mu=2 \,\mathrm{GeV}$ to give,
\bee
\sum_{q=u,d,s} \frac{F_{P}^{q/p}(0)}{m_q} \approx 6.43.
\ene

\subsection{Vector Current}

For the vector current, the matrix elements at the hadron
level are parameterized by two sets of 
form factors \(F_{1}^{q / N}\left(q^{2}\right)\) and \(F_{2}^{q / N}\left(q^{2}\right)\). 
For matching at the leading order, only their values evaluated at $q^2=0$ are necessary.
In the zero momentum transfer limit, the vector currents just count 
the number of valence quarks in the nucleon,
\begin{equation}
F_{1}^{u / p}(0)=2, \quad F_{1}^{d / p}(0)=1, \quad F_{1}^{s / p}(0)=0 \,.
\end{equation}
The form factors \(F_{2}^{q / N}(0)\) in
\geqn{eq:match:v} describe the quark contributions 
to the nucleon anomalous magnetic moments,
\begin{subequations}
\begin{align} 
  a_{p} 
& =
\frac{2}{3} F_{2}^{u / p}(0)-\frac{1}{3} F_{2}^{d / p}(0)-\frac{1}{3} F_{2}^{s / p}(0) 
\approx 
1.793 \,,
\\
  a_{n}
& =
\frac{2}{3} F_{2}^{u / n}(0)-\frac{1}{3} F_{2}^{d / n}(0)-\frac{1}{3} F_{2}^{s / n}(0) 
\approx
-1.913 \,.
\end{align}
\end{subequations}
The strange magnetic moment is given by \cite{Green:2015wqa,Sufian:2016pex},
\begin{equation}
F_{2}^{s / p}(0)=-0.064(17) \,.
\end{equation}
Then one can obtain,
\bee
F_{2}^{u / p}(0) = 1.609(17)\,,\quad F_{2}^{d / p}(0) = -2.097(17)\,.
\ene
Using these results we have,
\begin{equation}
  \sum_{q=u,d,s} F_{1}^{q/p}(0) = 3,
\qquad
  \sum_{q=u,d,s} F_{2}^{q/p}(0) = - 0.552.
\end{equation}

\subsection{Axial-Vector Current}

For the axial-vector operator, the hadron-level matrix elements 
are parametrized by two sets of form factors, $F_{A}^{q / N}\left(q^{2}\right)$ 
and $F_{P^{\prime}}^{q / N}\left(q^{2}\right)$.
At the leading order, 
only \(F_{A}^{q / N}(0)\) and the light meson pole contribution 
\(F_{P^{\prime}}^{q / N}\left(q^{2}\right)\) are necessary for our calculations.
The axial-vector form factor at zero momentum transfer can be calculated
from the matrix elements parameterization 
$2m_ps^\mu\Delta q_\mu= \langle p|\bar q \gamma_\mu \gamma_5 q | p\rangle_Q$
at scale $Q$ with $s^\mu$ being the proton polarization vector,
\begin{equation}
  F_{A}^{q / p}(0)
=
  \Delta q_{p} \,.
\end{equation}
For the light meson parts, it can be expanded as,
\begin{equation}
F_{P^{\prime}}^{q / N}\left(q^{2}\right)=\frac{m_{N}^{2}}{m_{\pi}^{2}-q^{2}} a_{P^{\prime}, \pi}^{q / N}+\frac{m_{N}^{2}}{m_{\eta}^{2}-q^{2}} a_{P^{\prime}, \eta}^{q / N}+\cdots\,.
\end{equation}
And the residua of the pion- and eta-pole contributions to \(F_{P^{\prime}}^{q / N}\) are given as,
\begin{subequations}
\begin{align}
  a_{P^{\prime}, \pi}^{u / p}
& =
-a_{P^{\prime}, \pi}^{d / p} 
=
2 g_{A} 
\,, \quad 
a_{P^{\prime}, \pi}^{s / p}=0 \,,
\\
  a_{P^{\prime}, \eta}^{u / p}
& =
a_{P^{\prime}, \eta}^{d / p}
=
-\frac{1}{2} a_{P^{\prime}, \eta}^{s / p}
=
\frac{2}{3}\left(\Delta u_{p}+\Delta d_{p}-2 \Delta s_{p}\right) \,.
\end{align}
\end{subequations}
Using these results we have,
\begin{equation}
  \sum_{q=u,d,s} F_{A}^{q/p}(0)
= 0.49 \,,
\qquad
  \sum_{q=u,d,s} F_{P'}^{q/p}(0)
\approx 0\,.
\end{equation}

\subsection{Tensor Operator}

The hadron-level matrix elements of the tensor operator are parameterized by three form factors, 
\(F_{T, 0}^{q / N}\left(q^{2}\right)\),
\(F_{T, 1}^{q / N}\left(q^{2}\right)\),
and \(F_{T, 2}^{q / N}\left(q^{2}\right)\).
They are related to the generalized tensor form factors
\cite{Gockeler:2005cj,Diehl:2001pm},
\begin{equation}
\hspace{-2mm}
  F_{T, 0}^{q / N}\left(q^{2}\right)
=
  m_{q} A_{T, 10}^{q / N}\left(q^{2}\right),
\quad
  F_{T, 1}^{q / N}\left(q^{2}\right)
=
-m_{q} B_{T, 10}^{q / N}\left(q^{2}\right),
\quad
  F_{T, 2}^{q / N}\left(q^{2}\right)
=
  \frac {m_{q}} 2 \tilde{A}_{T, 10}^{q / N}\left(q^{2}\right).
\end{equation}
At the leading order, only \(F_{T, 0}^{q / N}(0)\)
and \(F_{T, 1}^{q / N}(0)\) appear. 
The value of \(F_{T, 0}^{q / N}(0)\) is quite well determined
and
is usually expressed in terms of \(A_{T, 10}^{q / p}(0) \equiv g_{T}^{q}\) (with \(A_{T, 10}^{u(d) / p}=\)
\(A_{T, 10}^{d(u) / n}\) and \(A_{T, 10}^{s / p}=A_{T, 10}^{s / n}\) in the isospin limit).
The tensor charges $g_{T}^{q}$ are related to the transversity structure functions 
$\delta q_{N}(x, \mu)\) by \(g_{T}^{q}(\mu)=\) \(\int_{-1}^{1} d x \delta q_{N}(x, \mu)$.
The lattice calculations in the \(\overline{\mathrm{MS}}\) scheme at \(\mu=2\,\mathrm{GeV}\),
including both connected and disconnected contributions gives  
\cite{Bhattacharya:2016zcn,Alexandrou:2017qyt},
\begin{equation}
  g_{T}^{u}=0.794 \pm 0.015,
\quad
  g_{T}^{d}=-0.204 \pm 0.008,
\quad
  g_{T}^{s}=(3.2 \pm 8.6) \times 10^{-4}
\end{equation}
The other two form factors at zero momentum transfer, 
\(F_{T, 1}^{q / N}(0)\) and \(F_{T, 2}^{q / N}(0)\), 
are less well determined. The constituent quark model gives \cite{Pasquini:2005dk},
\begin{equation}
  B_{T, 10}^{u / p}(0) \approx 3.0\,,
\quad
  \tilde{A}_{T, 10}^{u / p} \approx-0.50 \,,
\quad 
  B_{T, 10}^{d / p}(0) \approx 0.24\,,
\quad
  \tilde{A}_{T, 10}^{d / p} \approx 0.46\,.
\end{equation}
For the $s$-quark component, we have following rough estimates,
\begin{equation}
- 0.2 \lesssim B_{T, 10}^{s / p}(0),
\quad
  \tilde{A}_{T, 10}^{s / p}(0) \lesssim 0.2
\end{equation}
We find the form factors are insensitive to the values of $B_{T, 10}^{s / p}(0)$
and $ \tilde{A}_{T, 10}^{s / p}(0)$.
Using these results we have,
\begin{equation*}
  \sum_{q=u,d,s} \frac{ F_{T,0}^{q/p}(0) }{m_q}
\approx 0.59,
\quad
  \sum_{q=u,d,s} \frac{ F_{T,1}^{q/p}(0) }{m_q}
\in \big[-2.96,\, - 2.56 \big]\,,
\quad
\sum_{q=u,d,s} \frac{ F_{T,2}^{q/p}(0) }{m_q}
\in \big[ - 0.12,\, 0.08\big]\,,
\end{equation*}
where the boundaries correspond to the lower and upper limits of 
$B_{T, 10}^{s / p}(0)$ and $ \tilde{A}_{T, 10}^{s / p}(0)$.
Our calculations use the lower boundaries to estimate the 
nucleon-level cross sections.

\section{Non-Relativistic Expansion}
\label{app:NR}

The non-relativistic expansion of the usual DM-nucleon interaction operators can be found
in \cite{DelNobile:2021wmp} and references within.
Since neutrino is always relativistic in our case,
we revisit the the non-relativistic expansions of both
the nucleon pair and DM-neutrino pair bilinears.
For completeness, we give our conventions for the spinor wave functions
in the Dirac representation as well as their non-relativistic expansions \cite{Hagiwara:2016rdv}.
In the Dirac representation, the Dirac matrixes are,
\bea
\gamma^{0}_{D} 
\equiv
\left(\begin{array}{cc}
1 & 0  
\\
0 & -1  
\end{array}\right)\,,\;\;\;
\vec{\gamma}_{D} 
\equiv
\left(\begin{array}{cc}
0 & \vec{\sigma}  
\\
-\vec{\sigma} & 0  
\end{array}\right)\,,\;\;\;
\gamma^{5}_{D} 
\equiv
\left(\begin{array}{cc}
0 & 1 
\\
1 & 0
\end{array}\right)\,.
\ena
The free particle solutions of the Dirac equation then
has large and small components,
\bea
  \frac{u_{D}( \vec{p}_{1}, s )}{ \sqrt{E+ m} }
=
\left(\begin{array}{c}
\xi_{s}
\\[3mm]
\dfrac{\vec{\sigma} \cdot \vec{p}_{1} }{E+m}\xi_{s}
\end{array}\right)
\,,\;\;\;
\frac{v_{D}( \vec{p}_{2}, r ) }{ \sqrt{E+ m} } =
\left(\begin{array}{c}
r \dfrac{\vec{\sigma} \cdot \vec{p}_{2} }{E+m}\xi_{-r}
\\[3mm]
r\xi_{-r}
\end{array}\right)\,,
\ena
and
\bea
\frac{\overline{u_{D}}( \vec{p}_{1}, s )}{ \sqrt{E+ m} }
=
\left(\begin{array}{cc}
\xi_{s}^\dag\quad\quad
&
\xi_{s}^\dag\dfrac{- \vec{\sigma} \cdot \vec{p}_{1} }{E+m}
\end{array}\right)
\,,\;\;\;
\frac{\overline{v_{D}}( \vec{p}_{2}, r )}{ \sqrt{E+ m} } =
\left(\begin{array}{cc}
r \xi_{-r}^\dag \dfrac{\vec{\sigma} \cdot \vec{p}_{2} }{E+m}\quad\quad
-r\xi_{-r}^\dag
\end{array}\right)\,,
\ena
where $\xi_{s}$ are eigenstates of the helicity operators
$\vec{\sigma} \cdot \bm p /|\bm p|$ with eigenvalue $s=\pm1$.
The two-component spinors can be explicitly written in terms
of the zenith ($\theta$) and azimuthal ($\phi$) angles,
\bea
  \xi_{+}
\equiv
\left(\begin{array}{c}
\cos(\theta/2)
\\
e^{i\phi}\sin(\theta/2)
\end{array}
\right),
\qquad
  \xi_{-}
\equiv 
\left(\begin{array}{c}
-e^{-i\phi}\sin(\theta/2)
\\
\cos(\theta/2)
\end{array}
\right).
\label{eq:xi}
\ena
We can easily check that
$\sum_{s=\pm} \xi_{s}\xi_{s}^\dag = \mathbb{I}$.
The spinor wave functions and the Dirac gamma matrices in the Dirac
representation are related to the ones in the chiral representation by
the following unitary transformation,
\bea
\psi_{D} \equiv  U_{D} \psi U_{D}^{-1}
\,,\;\;\;
\gamma^{\mu}_{D} \equiv U_{D} \gamma_{C}^{\mu}U_{D}^{-1}
\,,\;\;\;
U_{D} \equiv \frac{1}{\sqrt[]{2}}
\left(\begin{array}{cc}
1 & 1
\\
-1 & 1
\end{array}\right)\,.
\ena

Using these wave functions, we can easily find the non-relativistic 
expansions of the nucleon bilinears,
\begin{subequations}
\begin{align}
  \overline{u}^{N'}_{h_{N'}} u^{N}_{h_{N}} 
& \stackrel{\mathrm{NR}}{\approx}
2 m_{\mathrm{N}} \, (\omega_{h_{N'}}^{N'})^\dag \omega_{h_{N}}^{N}, 
\\
  \overline{u}^{N'}_{h_{N'}} ( i \gamma^5 ) u^{N}_{h_{N}} 
& \stackrel{\mathrm{NR}}{ \approx}
- i (\omega_{h_{N'}}^{N'})^\dag \big( \boldsymbol{q} \cdot \boldsymbol{\sigma} \big) 
\omega_{h_{N}}^{N}, 
\\
  \overline{u}^{N'}_{h_{N'}} \gamma^\mu u^{N}_{h_{N}} 
& \stackrel{\mathrm{NR}}{ \approx}
(\omega_{h_{N'}}^{N'})^\dag
\left
[\begin{array}{cc}
  2 m_{\mathrm{N}} \mathbb{I},
& \quad
  \boldsymbol{p}\mathbb{I} 
- i \boldsymbol{q} \times \boldsymbol{\sigma}
\end{array}
\right] \omega_{h_{N}}^{N} \,, 
\\
  \overline{u}^{N'}_{h_{N'}} \gamma^\mu\gamma_5 u^{N}_{h_{N}} 
& \stackrel{\mathrm{NR}}{ \approx}
(\omega_{h_{N'}}^{N'})^\dag
\left[
\begin{array}{cc}
  \boldsymbol{p} \cdot \boldsymbol{\sigma} \,,
& \quad
  2 m_{\mathrm{N}}  \boldsymbol{\sigma}
\end{array}
\right] \omega_{h_{N}}^{N}\,, 
\\
  \overline{u}^{N'}_{h_{N'}} \sigma^{\mu \nu}   u^{N}_{h_{N}} 
& \stackrel{\mathrm{NR}}{ \approx}
(\omega_{h_{N'}}^{N'})^\dag 
\left[
\begin{array}{cc}
0 &\quad i \boldsymbol{q}\mathbb{I}   + \boldsymbol{p} \times \boldsymbol{\sigma} 
\\
-i \boldsymbol{q}\mathbb{I}  - \boldsymbol{p} \times \boldsymbol{\sigma}   &\quad
2m_{\rm N} \varepsilon_{i j k}  \boldsymbol{\sigma}_{k} 
\end{array}
\right] \omega_{h_{N}}^{N} ,
\end{align}
\end{subequations}
where we have used the abbreviated notations 
$u^{N}_{h_{N}} \equiv u_N(\boldsymbol{p}_i, h_{N})$ 
and $\overline{u}^{N'}_{h_{N'}} \equiv \overline{u}_{N'}(\boldsymbol{p}_f, h_{N'})$
while $\omega$ is the nucleon spin vectors as
defined in \geqn{eq:xi}.
The total and relative momentum are defined as 
$\boldsymbol{p} \equiv \boldsymbol{p}_f + \boldsymbol{p}_i$ and
$\boldsymbol{q} \equiv \boldsymbol{p}_f - \boldsymbol{p}_i$, respectively.
Similarly for the DM-neutrino system, the corresponding wave functions are given as,
\bee
\frac{\overline{u_{D}}( \vec{p}_{\nu}, h_\nu )}{ \sqrt{E_\nu} } =
\left(
\begin{array}{cc}
\xi_{h_\nu}^\dag
&
- h_\nu\, \xi_{h_\nu}^\dag
\end{array}
\right),
\qquad
  \frac{u_{D}( \vec{p}_{\chi}, h_\chi )}{ \sqrt{2m_\chi} } =
\left(
\begin{array}{c}
\xi_{h_\chi}
\\
\dfrac{\vec{\sigma} \cdot \vec{p}_{\chi} }{2m_\chi}\xi_{h_\chi}
\end{array}
\right)\,.
\ene
Please note that the neutrino can only be left-handed, \ie, $h_\nu = -1$.
After similar straightforward calculations, one can find the following
non-relativistic expressions of various bilinears,
\begin{subequations}
\begin{align}
  \overline{u}^\nu_{h_\nu} u^\chi_{h_\chi} 
& \stackrel{\mathrm{NR}}{ \approx}
  \sqrt{ 2 m_{\chi} E_\nu } (\xi_{h_\nu}^\nu)^\dag \xi_{h_\chi}^\chi, 
\\
  \overline{u}^\nu_{h_\nu} ( i \gamma^5 ) u^\chi_{h_\chi} 
& \stackrel{\mathrm{NR}}{ \approx}
- i s_2 \sqrt{ 2 m_{\chi} E_\nu } (\xi_{h_\nu}^\nu)^\dag \xi_{h_\chi}^\chi, 
\\
  \overline{u}^\nu_{h_\nu} \gamma^\mu  u^\chi_{h_\chi} 
& \stackrel{\mathrm{NR}}{ \approx}
  \sqrt{ 2 m_{\chi} E_\nu } (\xi_{h_\nu}^\nu)^\dag
\left[
\begin{array}{cc}
\mathbb{I}  \,, &\quad
h_\nu  \boldsymbol{\sigma} 
\end{array}
\right] \xi_{h_\chi}^\chi , 
\\
  \overline{u}^\nu_{h_\nu} \gamma^\mu\gamma_5  u^\chi_{h_\chi} 
& \stackrel{\mathrm{NR}}{ \approx}
\sqrt{ 2 m_{\chi} E_\nu } (\xi_{h_\nu}^\nu)^\dag
\left[
\begin{array}{cc}
  h_\nu \mathbb{I} \,,
& \quad \boldsymbol{\sigma} 
\end{array}
\right]\xi_{h_\chi}^\chi , 
\\
  \overline{u}^\nu_{h_\nu} \sigma^{\mu \nu}  u^\chi_{h_\chi} 
& \stackrel{\mathrm{NR}}{ \approx}
\sqrt{ 2 m_{\chi} E_\nu } (\xi_{h_\nu}^\nu)^\dag
\left[
\begin{array}{cc}
0 &\quad - i h_\nu \boldsymbol{\sigma}  
\\
 i h_\nu \boldsymbol{\sigma}   &\quad -  \boldsymbol{\sigma} 
\end{array}
\right] \xi_{h_\chi}^\chi ,
\end{align}
\end{subequations}
where $\overline{u}^\nu_{h_\nu} \equiv \overline{u_\nu}(\vec{p}_\nu, h_\nu)$
and $u^\chi_{h_\chi} \equiv u_\chi(\vec{p}_\chi, h_\chi) $.

\providecommand{\href}[2]{#2}\begingroup\raggedright

\end{document}